\documentclass[11pt, a4paper, oneside]{Thesis} 

\graphicspath{{./Pictures/}} 
\usepackage{textcomp}
\usepackage{multirow}
\usepackage{multicol}
\usepackage{subfiles}
\usepackage{booktabs}
\usepackage{tabularx}
\usepackage{lscape}
\usepackage{hyperref}
\usepackage{tabularray}
\usepackage{enumitem}

\usepackage[square, numbers, comma, sort & compress]{natbib} 
\hypersetup{urlcolor=blue, colorlinks=true} 
\title{\ttitle} 

\begin{document}

\frontmatter 

\setstretch{1.3} 

\fancyhead{} 
\rhead{\thepage} 
\lhead{} 

\pagestyle{fancy} 

\newcommand{\HRule}{\rule{\linewidth}{0.5mm}} 

\providecommand{\keywords}[1]
{
  \small	
  \textbf{\textit{Keywords---}} #1
}

\hypersetup{pdftitle={\ttitle}}
\hypersetup{pdfsubject=\subjectname}
\hypersetup{pdfauthor=\authornames}
\hypersetup{pdfkeywords=\keywordnames}


\begin{titlepage}
\begin{center}
\begin{figure}[h]
\centering
\includegraphics[width=0.12\textwidth]{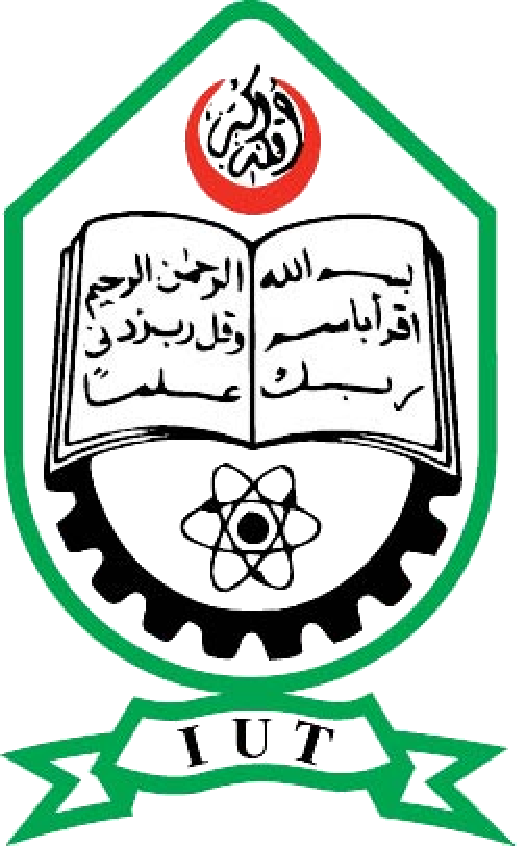}
\end{figure}

\textsc{\LARGE ISLAMIC UNIVERSITY OF TECHNOLOGY}\\[1cm] 

\HRule \\[0.4cm] 
{\LARGE \bfseries \ttitle}\\[0.4cm] 
\HRule \\[1cm] 

\emph{Authors }\\
\textbf{Md. Maksudul Haque (180041222)}\\
\textbf{Samiul Islam  (180041221)}\\
\textbf{Abu Jobayer Md. Sadat (180041202)}\\[1cm]

\begin{center}
    \textbf{Supervised By}\\
    \vspace{2mm}
    \begin{table}[h]
    \centering
    \begin{spacing}{1}
    \begin{tabular}{ccc}
    \begin{tabular}[c]{@{}c@{}}Dr. Hasan Mahmud\\ Associate Professor\end{tabular} \quad \quad \quad & \begin{tabular}[c]{@{}c@{}}\hspace{5pt}Fardin Saad\\ \hspace{5pt}Lecturer\end{tabular} \quad \quad \quad & \begin{tabular}[c]{@{}c@{}}Dr. Md. Kamrul Hasan \\ Professor\end{tabular}
    \end{tabular}
    \end{spacing}
    \end{table}
    
    \vspace{6mm}

    \begin{spacing}{1}
    Systems and Software Lab (SSL)\\
    Department of Computer Science and Engineering (CSE)\\
    Islamic University of Technology (IUT)\\
    A subsidiary organ of the Organization of Islamic Cooperation (OIC)
      \end{spacing}
\end{center}

\vspace{5mm}
\begin{center}
\begin{spacing}{1}
    \small
    \textit{A thesis submitted to the Department of CSE\\
    in partial fulfillment of the requirements for the degree of B.Sc. Engineering in Computer Science and Engineering (CSE)\\
    \textbf{Academic Year: 2021-2022}}\\
    \textbf{\today}
\end{spacing}
    
\end{center}

    
    

\end{center}
\end{titlepage}


\Declaration{


    This is to certify that the work presented in this thesis titled \textbf{“Capturing Spectral and Long-term Contextual
    Information for Speech Emotion Recognition Using Deep Learning Techniques"} is the outcome of the analysis and experiments carried out by Md. Maksudul Haque, Samiul Islam and Abu Jobayer Md. Sadat under the supervision of the Systems and Software Lab (SSL) group of the Department of Computer Science and Engineering (CSE), Islamic University of Technology (IUT), Dhaka, Bangladesh. It is also declared that neither this thesis nor any part of this thesis has been submitted anywhere else for any degree or diploma. Information derived from the published and unpublished work of others has been acknowledged in the text and a list of references is given.
 
 }
Authors:\\ [1.2cm]
\rule[.5em]{25em}{0.1pt}\\
\textbf{Md. Maksudul Haque}\\
Student ID - \textbf{180041222}\\

\rule[.5em]{25em}{0.1pt}\\
\textbf{Samiul Islam}\\
Student ID - \textbf{180041221}\\

\rule[.5em]{25em}{0.1pt}\\
\textbf{Abu Jobayer Md. Sadat}\\
Student ID - \textbf{180041202}\\

\clearpage 

\addtotoc{Approval}


\large Supervisors:\\[1.5cm]

\rule[.5em]{25em}{0.5pt}\\
\textbf{Dr. Hasan Mahmud} \\
Associate Professor\\
Systems and Software Lab (SSL)\\
Department of Computer Science and Engineering (CSE)\\
Islamic University of Technology (IUT)\\[1.9cm] 

\rule[.5em]{25em}{0.5pt}\\
\textbf{Fardin Saad}\\
Lecturer,\\
Systems and Software Lab (SSL)\\
Department of Computer Science and Engineering (CSE)\\
Islamic University of Technology (IUT)\\[1.9cm]

\rule[.5em]{25em}{0.5pt}\\
\textbf{Dr. Md. Kamrul Hasan} \\
Professor,\\
Systems and Software Lab (SSL)\\
Department of Computer Science and Engineering (CSE)\\
Islamic University of Technology (IUT)\\[1.9cm]

\clearpage 


\addtotoc{Abstract} 

\abstract{\addtocontents{toc}{\vspace{1em}} 


 Traditional approaches in speech emotion recognition, such as LSTM, CNN, RNN, SVM, and MLP, have limitations such as difficulty capturing long-term dependencies in sequential data, capturing the temporal dynamics, and struggling to capture complex patterns and relationships in multimodal data. This research addresses these shortcomings by proposing an ensemble model that combines Graph Convolutional Networks (GCN) for processing textual data and the HuBERT transformer for analyzing audio signals. We found that GCNs excel at capturing Long-term contextual dependencies and relationships within textual data by leveraging graph-based representations of text and thus detecting the contextual meaning and semantic relationships between words. On the other hand, HuBERT utilizes self-attention mechanisms to capture long-range dependencies, enabling the modeling of temporal dynamics present in speech and capturing subtle nuances and variations that contribute to emotion recognition. By combining GCN and HuBERT, our ensemble model can leverage the strengths of both approaches. This allows for the simultaneous analysis of multimodal data, and the fusion of these modalities enables the extraction of complementary information, enhancing the discriminative power of the emotion recognition system. The results indicate that the combined model can overcome the limitations of traditional methods, leading to enhanced accuracy in recognizing emotions from speech.
\newline
}


\clearpage 


\setstretch{1.3} 

\acknowledgements{\addtocontents{toc}{\vspace{1em}} 
 We would like to express our grateful appreciation to our research group \textbf{Systems and Software Lab (SSL)} and our supervisor \textbf{Associate Professor Dr. Hasan Mahmud, Professor Dr. Md. Kamrul Hasan, and Lecturer Fardin Saad,} Department of Computer Science \& Engineering, IUT for being our advisers and mentors. Their motivation, suggestions, and insights for this research have been invaluable. Without their support and proper guidance, this research would never have been possible. Their valuable opinion, time, and input were provided throughout the thesis work, from the first phase of thesis topics introduction, subject selection, proposing algorithm, and modification till the project implementation and finalization which helped us to do our thesis work in the proper way. We are really grateful to them.


}

\clearpage 


\pagestyle{fancy} 

\lhead{\emph{Contents}} 
\tableofcontents 

\lhead{\emph{List of Figures}} 
\listoffigures 

\lhead{\emph{List of Tables}} 
\listoftables 

\clearpage


%
%
%


\setstretch{1.3} 

\pagestyle{empty} 

\dedicatory{Dedicated to our parents, mentors, and teachers for their continuous encouragement of our academic endeavors and research \ldots} 

\addtocontents{toc}{\vspace{2em}} 


\mainmatter 

\pagestyle{fancy} 

\chapter{Introduction} 
\label{Introduction} 
\lhead{Chapter \ref{Introduction}. \emph{Introduction}} 

   \justify
    The task of automatically detecting the emotional content of spoken language is defined as speech emotion recognition (SER). The naturalness and human-likeness of virtual assistants can be improved, the user experience of social robots can be improved, and people with communication difficulties can get feedback courtesy of this. SER can be difficult for a variety of reasons. First of all, emotions are illogical and might differ from one individual to the next. Second, the same feeling may be communicated in a variety of ways based on the speaker's temperament, cultural background, and conversational context. Third, voice signals may be distorted and include extraneous data, such as background noise or the accent of the speaker. Several techniques have been placed forth for SER, including more established deep learning methods like convolutional neural networks (CNNs) and long short-term memory (LSTM) networks, as well as more established machine learning methods like support vector machines (SVMs) and hidden Markov models (HMMs). The precise properties of the data and the task's needs determine the SER strategy to use. In general, more complex models, like deep learning techniques, tend to be more accurate but may need more computer power and labeled data in larger amounts. Overall, SER is a dynamic field of study with a variety of difficulties and chances for growth and progress.

    \justify
    Recognition of emotion from the human speech is a non-trivial process for machines. Humans can understand each other's feelings through speech very easily while machines it goes through a computational procedure. Deep learning approaches for automated speech emotion recognition have made it easier for people to interact with computers supporting the Human-Computer Interaction field. There are a few techniques to extract emotional features from speech, including extraction of hand-crafted prosodic features and automatic feature learning methods. Automated feature-learning algorithms learn relevant features from the input data. Once important features are extracted, then it will be fed into the machine learning architecture to classify emotions accordingly.
    
    \justify
    Speech emotion recognition involves analyzing various acoustic and linguistic features present in the speech signal to discern emotions such as sadness, anger, happiness, neutrality, surprise, fear, and disgust. By capturing vocal cues, intonations, prosodic patterns, and language characteristics, speech emotion recognition aims to decode the underlying emotional information and provide valuable insights into human communication. This field has significant implications in diverse domains, including human-computer interaction, sentiment analysis, mental health monitoring, and personalized user experiences, where understanding and responding to human emotions are essential for effective communication and interaction.

    \justify
    Speech emotion recognition holds immense importance as it allows for a deeper understanding and analysis of human emotions conveyed through spoken language. Emotions play a pivotal role in communication, shaping the way messages are delivered, perceived, and interpreted. By accurately recognizing and comprehending emotions from speech, a wide range of applications and fields can benefit greatly. In the domain of human-computer interaction, speech-emotion recognition enables more intuitive and natural interactions between humans and machines. By discerning the emotional states of users, virtual assistants, and chatbots can respond in a more personalized and empathetic manner, enhancing user experiences and satisfaction. This capability has the potential to revolutionize the way we interact with technology, making it more human-centric and emotionally aware. Furthermore, in the realm of sentiment analysis, speech emotion recognition allows for the automatic assessment of emotions expressed in audio content, such as customer reviews, social media posts, and recorded interviews. This enables businesses and organizations to gain insights into customer sentiment, evaluate brand perception, and tailor their marketing strategies accordingly. Sentiment analysis powered by speech emotion recognition also aids in understanding public opinion, tracking trends, and informing decision-making processes. Additionally, speech-emotion recognition plays a crucial role in mental health monitoring and assessment. By analyzing speech patterns and emotional cues, this technology can assist in the early detection of mental health disorders such as depression, anxiety, and bipolar disorder. It provides a non-intrusive and objective means of evaluating an individual's emotional well-being, facilitating timely interventions, and improving mental healthcare outcomes.

    \justify
    Recent developments in speech emotion recognition have focused on leveraging advanced machine learning techniques and multimodal approaches. Researchers have explored the use of deep learning models, including recurrent neural networks (RNNs), convolutional neural networks (CNNs), and transformers, to capture intricate patterns and temporal dependencies in speech data. Multimodal approaches have gained attention, combining audio, textual, and visual cues to enhance emotion recognition performance.

    \justify
    The challenge of speech emotion recognition lies in determining which speech features hold the most significant discriminatory power in distinguishing between these emotions. Traditionally, speech emotion recognition comprises two main stages: feature extraction and classification. During the feature extraction stage, the speech signal is transformed into numerical values using a range of front-end signal processing techniques. Once the features are extracted, the classification stage involves assigning the appropriate emotional label to the speaker's utterances. Machine learning algorithms, such as support vector machines (SVMs), hidden Markov models (HMMs), and deep learning models, are commonly employed for classification. These algorithms learn from the extracted features and develop decision boundaries to accurately classify speech into the corresponding emotional categories.

    \justify
    Speech emotion recognition is a challenging task due to the intricacies involved in understanding and interpreting emotional cues from speech. Identifying the most powerful features and developing robust classification models are ongoing areas of research. In this report, we propose a novel approach that combines Graph Convolutional Networks (GCN) and the HuBERT transformer model to enhance feature extraction from both textual and audio modalities. By leveraging the strengths of these models, we aim to overcome the limitations of traditional methods and achieve improved accuracy in recognizing emotions from speech.

\section{Problem Statement}
    As speech signals contain various acoustic and linguistic cues, combining both text and speech modalities provides a more comprehensive and rich representation of emotional cues. Efficiently capturing acoustic features which contain valuable emotional cues from speech modality and semantic and contextual information from textual modality is a difficult task. The traditional approach of feature extraction and classification does not effectively capture long-term dependencies in sequential data and the temporal dynamics effectively and struggles to capture and integrate complex patterns and relationships.  To overcome these challenges, a novel multimodal approach combining Graph Convolutional Networks (GCN) and the HuBERT transformer is proposed to enhance feature extraction and improve accuracy in speech emotion recognition.

\section{Motivation}    
\justify
 The motivation behind this project is driven by the need to improve the accuracy and robustness of speech-emotion recognition systems. Emotions play a fundamental role in human communication and accurately recognizing and understanding them from speech can have significant implications in various domains. However, traditional approaches in speech emotion recognition have limitations in effectively capturing and integrating information from multimodal sources. There is a growing demand for novel techniques that can leverage both textual and audio modalities to enhance feature extraction and overcome the shortcomings of existing methods. By addressing these challenges, this project aims to advance the field of speech emotion recognition and contribute to the development of more sophisticated and reliable emotion recognition systems.\\
 
    The scope of this project encompasses the development and evaluation of a novel approach for multimodal speech emotion recognition using Graph Convolutional Networks (GCN) and the HuBERT transformer. The project involves preprocessing and preparing two datasets, namely the IEMOCAP and RAVDESS datasets, for experimentation. The audio signals from these datasets will be transformed into text, and both the textual and audio data will be processed using the GCN and HuBERT models. The test results from these models will be stored and analyzed. The project aims to explore the benefits of combining GCN and HuBERT in capturing informative features and information from both speech and text modalities. The proposed approach will be compared against traditional methods, such as LSTM, CNN, RNN, SVM, and MLP, to demonstrate its superiority in speech emotion recognition. The project's scope also includes evaluating the performance of the approach in terms of accuracy, efficiency, and ability to handle multimodal data. Overall, the project seeks to contribute to the advancement of multimodal speech emotion recognition techniques and provide insights for further research in this domain.

\pagebreak
\section{Research Challenges}
\justify
    Preprocessing the IEMOCAP and RAVDESS datasets posed specific challenges. In the case of IEMOCAP, the dataset contained instances of mixed emotions within each speech sample, making it difficult to create accurate emotion graphs for the GCN model. Resolving this challenge required careful consideration of the annotation and representation of mixed emotions to ensure the effective integration of the data into the research framework. Additionally, both datasets required preprocessing steps to handle data inconsistencies, noise reduction, and normalization to improve the overall data quality and reduce unwanted variations.

    \justify
    Extracting informative and discriminative features from speech data posed a significant research challenge. Various acoustic, prosodic, and linguistic features contribute to the expression of emotions in speech. However, automatically extracting these features and representing them effectively for emotion recognition required careful consideration of feature selection, dimensionality reduction techniques, and appropriate modeling approaches. The challenge lay in identifying the most relevant features and developing automated methods to extract and represent them accurately, considering the dynamic nature of emotions and the variability across different speakers and utterances.

    \justify
    We encountered a research gap regarding the application of the HuBERT transformer specifically for speech emotion recognition. While there have been advancements in transformer models for natural language processing tasks, their utilization in the domain of speech emotion recognition was limited. This lack of prior work presented a challenge in terms of adapting the HuBERT model and exploring its effectiveness in capturing the emotional features present in speech data. Extensive experimentation and fine-tuning were necessary to determine the optimal configuration and performance of HuBERT for this task.

\section{Research Contribution}
 \justify
    Novel Multimodal Approach: Our thesis contributes a novel multimodal approach for speech emotion recognition by combining the HuBERT model for spectral representation of speech data and the GCN model for textual representation of speech data. This combination has not been explored in previous studies.

    \justify
    Enhanced Model Performance: We found that our multimodal model outperforms other existing multimodal models by utilizing GCN for capturing long-term contextual dependencies and semantic relationships in text. GCNs leverage graph-based representations, which are effective in capturing the nuanced expressions of emotions in textual data. Additionally, the use of HuBERT, which is specifically designed for speech-related tasks, improves the model's performance compared to other transformer models (e.g., stack transformer, traditional BERT) in the speech emotion recognition task.

    \justify
    HuBERT for Speech Emotion Recognition: We contribute to the field by applying the HuBERT model, which was previously used for automatic speech recognition, to the speech emotion recognition task. This expands the application of HuBERT and demonstrates its effectiveness in capturing emotional cues from speech data.

    \justify
    Evaluation on Benchmark Datasets: Our thesis includes the evaluation of the proposed models on the latest benchmark datasets, namely the RAVDESS and IEMOCAP datasets. By testing our models on these widely used datasets, we provide valuable insights into their performance and generalizability.

    \justify
    GCN and HuBERT for Speech Emotion Recognition: As previously expressed, we tried to compress our architecture in a parallel HuBERT and GCN module reasoning to capture both spectral and Long-term Contextual information of speech so that recognition is more robust and emotion can be classified depending on both the speech information at the same time. Also going for a multimodal approach is always a win-win situation for downstream tasks like this. Hence our motivation from the given challenges is like this to go for as our research contribution.

    \justify
    Overall, our thesis makes significant contributions by introducing a novel multimodal approach, highlighting the strengths of GCN and HuBERT models, extending the application of HuBERT to speech emotion recognition, and evaluating the models on relevant benchmark datasets. These contributions advance the field of speech emotion recognition and provide valuable insights for future research in multimodal emotion analysis.

\pagebreak
\section{Thesis Outline}
    \justify
    In Chapter \ref{Introduction} we have introduced the problem statement and research objectives and outline the significance and relevance of multimodal speech emotion recognition. In Chapter \ref{Background}, we have explored the concepts of different speech features such as Mel-frequency cepstral coefficients (MFCC), acoustic features, prosodic features, self-supervision mechanism, semantic relationships, contextual dependencies in textual representation, and long-term dependencies in sequential data. In Chapter \ref{Literature}, we have reviewed speech emotion recognition using traditional machine learning approaches such as Support Vector Machine, Hidden Markov Model, and Gaussian Mixture Model, as well as deep learning approaches including Generic Deep Neural Network, Deep autoencoder, Convolutional Neural Network, Recurrent Neural Network, and the evolution of deep learning techniques based on recent benchmark literature. In chapter \ref{methodology} we have presented Discusses the preparation of datasets, specifically IEMOCAP and RAVDESS, presents the skeleton of the first proposed method, and outlines the final implemented pipeline. Then we have covered the selection of baseline models, details the experiments conducted, describes the experiment setting and pre-processing steps, sets the hyper-parameters, and defines the evaluation metrics in Chapter \ref{experiment}. In Chapter \ref{result} we have presented and discussed the classification results of Experiment 1, Experiment 2, and Experiment 3.In the final Chapter \ref{Conclusion}, we have summarized the main findings of the thesis, draws conclusions, and suggests potential directions for future research.



\chapter{Background Study} 

\label{Background} 

\lhead{Chapter \ref{Background}. \emph{Background study}}

 \justify
    The background study for this project encompasses various aspects related to speech emotion recognition. Understanding and addressing these various aspects is crucial for the successful development of accurate and robust speech-emotion recognition systems. Speech emotion recognition tasks typically involve analyzing various acoustic and linguistic features to determine the emotional content of speech. Here are some commonly used speech features for this task.

\section{Features Used in Speech Emotion Recognition}
Speech Emotion Recognition (SER) involves extracting relevant features from speech signals to identify and classify different emotional states. Various types of features are utilized in SER systems. Acoustic features capture the characteristics of the speech signal, such as pitch, energy, and spectral properties. Prosodic features focus on aspects like intonation, rhythm, and speech rate, which convey emotional information. Additionally, contextual features take into account the linguistic context and semantic relationships within the speech. Furthermore, long-term dependencies in sequential data and contextual dependencies in textual representations are considered. These features collectively provide valuable information for accurate recognition and classification of emotions in speech, enabling SER systems to capture the nuanced variations in emotional expression.

\subsection{Mel-frequency cepstral coefficients (MFCC)}
    \justify
    The Mel-frequency spectrum, also known as Mel spectrogram or Mel-frequency cepstral coefficients (MFCC), is a representation of the speech signal that mimics the human auditory system's perception of different frequencies. It is commonly used in speech analysis tasks, including speech emotion recognition.

    \justify
    In speech emotion recognition, the Mel-frequency spectrum is important because it captures essential acoustic features related to emotions. Emotions are often expressed through changes in pitch, intensity, and timbre, which are reflected in the distribution of energy across different frequency bands. By analyzing the Mel-frequency spectrum, we can extract relevant features that characterize emotional expressions in speech.

    \subsection{Acoustic features}
    \justify
    Acoustic features in speech emotion recognition refer to the characteristics of the speech signal that are related to its acoustic properties. These features capture various aspects of the signal's waveform, such as pitch, intensity, duration, spectral content, and prosody. These features provide non-verbal cues, are often universally applicable across languages and cultures, robust to noise, and allow for real-time processing. By analyzing acoustic features, we can extract valuable information that contributes to accurate emotion recognition and enables the development of practical and efficient speech emotion.

    \subsection{Prosodic features}
    Prosodic features in speech emotion recognition refer to the acoustic characteristics that capture the rhythm, intonation, and stress patterns of speech. They include elements such as pitch contour, speech rate, pause duration, and emphasis on certain words or syllables. Prosodic features are important in speech emotion recognition because they convey crucial information about emotional expressiveness. Emotions are often reflected in the way speech is delivered, including changes in pitch, rhythm, and stress. By analyzing prosodic features, we can capture the melodic and rhythmic aspects of speech, which contribute to the expression and perception of emotions. Incorporating prosodic features enhances the accuracy and robustness of emotion recognition systems by considering the nuanced variations in speech patterns associated with different emotional states.

    \subsection{Self-supervision mechanism}
    \justify
    Self-supervised learning enables the model to learn high-level representations and features directly from the raw speech signal. By training the model on self-supervised tasks, such as predicting masked segments or reconstructing corrupted speech, the model can learn to extract discriminative and contextually rich features that are informative for emotion recognition. Moreover, by learning from unlabeled data, self-supervised models can capture general representations and patterns that are relevant to emotion recognition tasks. Also, self-supervised learning acts as a pre training phase, where the model learns general representations from unlabeled data. These learned representations can then be fine-tuned and applied to specific emotion recognition tasks with limited labeled data.

    \subsection{Semantic relationships}
    \justify
    Semantic relationships in speech emotion recognition refer to the connections and associations between words and their meanings within the context of emotional expression. It involves understanding the relationships and dependencies between words and the emotions they convey, as well as how they contribute to the overall emotional content of the speech. Semantic relationships play a vital role in speech emotion recognition by providing contextual understanding, enabling fine-grained emotion recognition, interpreting figurative language, disambiguating polysemous words, and facilitating multimodal integration. By capturing these relationships, the model can better comprehend and interpret the emotional content expressed in speech, leading to improved performance in emotion recognition tasks.

    \subsection{Contextual dependencies in textual representation of data}
    \justify
    In the textual representation of speech emotion recognition task, contextual dependencies refer to the relationships and dependencies between words or textual elements within a given context or sentence. These dependencies capture the influence and meaning conveyed by the surrounding words and help in understanding the emotional content.

    \justify
    Contextual dependencies are important in textual representation for speech emotion recognition for several reasons. Firstly, emotions are often expressed and perceived through the choice of words, sentence structure, and the overall context. The surrounding words provide contextual cues that contribute to the interpretation and understanding of emotions.

    \justify
    Secondly, capturing contextual dependencies allows for a more comprehensive analysis of the emotional content. By considering the relationships between words, such as their semantic connections, syntactic structure, and co-occurrence patterns, the model can gain a better understanding of the overall emotional expression.

    \justify
    Additionally, contextual dependencies aid in disambiguating the meaning of words. Words may have multiple meanings or interpretations, but their context helps in discerning the intended emotional connotation. By considering the surrounding words, the model can better infer the emotional intent and nuances in the textual representation.

    \subsection{Long-term dependencies in sequential data}
    \justify
    Long-term dependencies in sequential data refer to the relationships and dependencies that exist between distant elements or events in a sequence. In the context of speech emotion recognition, long-term dependencies are crucial for understanding the context and temporal dynamics of speech. They involve capturing how earlier segments of speech may influence the interpretation and understanding of subsequent segments.

    \justify
    Traditional approaches like LSTMs and RNNs have limitations in effectively modeling and capturing these long-term dependencies. These models suffer from vanishing or exploding gradients, making it difficult to propagate information across long sequences. As a result, they may struggle to capture the contextual meaning and relationships between words or phonemes that are further apart in the speech sequence.

    \justify
    To address this challenge, Graph Convolutional Networks (GCNs) offer a novel approach. GCNs leverage graph-based representations of text, where words or phonemes are represented as nodes in a graph, and the relationships between them are captured through edges. This graph representation allows GCNs to capture long-term contextual dependencies and semantic relationships more effectively. By considering the connections and interactions between different elements in the graph, GCNs can better model the dependencies in sequential data.

\section{Machine learning models used in Speech emotion Recognition}
The field of Speech Emotion Recognition (SER) has evolved significantly in terms of the models used for classification. Initially, traditional handcrafted feature-based models were employed, utilizing features such as pitch, energy, and spectral properties. However, with the advent of machine learning, more sophisticated approaches were developed, including Support Vector Machines (SVM), Hidden Markov Models (HMM), and Gaussian Mixture Models (GMM). These models leveraged the power of statistical learning algorithms to improve SER performance. As deep learning gained prominence, deep neural networks, such as Convolutional Neural Networks (CNN) and Recurrent Neural Networks (RNN), were introduced, offering enhanced representation learning capabilities. These models demonstrated superior performance by automatically extracting meaningful features from raw speech data. More recently, with the rise of language models, the SER task has incorporated the use of pre-trained language models, such as GPT (Generative Pre-trained Transformer) and BERT (Bidirectional Encoder Representations from Transformers), which can leverage contextual information and semantic relationships within the speech data, enabling more accurate and nuanced emotion recognition. This progression from traditional handcrafted features to machine learning models and now to language models showcases the continuous advancements in SER research, highlighting the impact of these models in capturing and understanding emotional expressions in speech.

\subsection{Traditional models}
Traditional models used in Speech Emotion Recognition (SER) encompass various techniques, including Support Vector Machines (SVM), Hidden Markov Models (HMM), and Gaussian Mixture Models (GMM). A brief overview of some of the traditional models is given below:
\begin{itemize}
    \item \textbf{SVM:} SVM is a supervised learning model that separates different emotion classes using a hyperplane in a high-dimensional feature space. Its strength lies in its ability to handle high-dimensional feature vectors and handle non-linear relationships. However, SVM's performance heavily relies on the choice of kernel function and hyperparameters, and it may struggle with imbalanced datasets.
    \item \textbf{HMM:} HMM is a statistical model that captures temporal dependencies in speech signals by modeling the underlying emotional state transitions. It can effectively capture sequential information and has been widely used in speech and audio processing. However, HMM assumes that the emotion classes are generated from discrete states, which may limit its ability to capture complex emotions.
    \item \textbf{GMM:} GMM is a probabilistic model that represents speech data as a mixture of Gaussian distributions. It is flexible, capable of capturing complex statistical relationships, and can model the variability within each emotion class. However, GMM may struggle with modeling long-term dependencies and can be computationally demanding.

\end{itemize} 

Over time, traditional models have evolved to incorporate advanced techniques and improvements. For example, SVM has been enhanced with kernel trick methods and ensemble learning. HMM has been combined with other models like GMM to create hybrid systems, and adaptations like the semi-continuous HMM have been developed. GMM models have seen advancements in the form of diagonal covariance matrices and the integration of feature transformations.

While traditional models have provided valuable insights and achieved reasonable performance in SER, their reliance on handcrafted features and limited ability to capture complex relationships led to the emergence of more powerful models, such as deep learning and language models, which offer improved representation learning and contextual understanding of emotions in speech.

\subsection{Deep learning models}
Deep learning models have overcome the limitations of traditional models in Speech Emotion Recognition (SER) by leveraging their ability to automatically learn complex features from raw speech data. They can capture both local and global dependencies, adapt to various data distributions, and handle high-dimensional input effectively. Some notable deep learning models used in SER include Convolutional Neural Networks (CNN), Recurrent Neural Networks (RNN), and their variants.

CNNs excel at extracting local patterns and features by utilizing convolutional filters. They can capture spatial and temporal information from spectrogram representations of speech signals, enabling effective feature extraction. However, CNNs may struggle with modeling long-term dependencies and may not fully capture the sequential nature of speech data.

RNNs, particularly the Long Short-Term Memory (LSTM) and Gated Recurrent Unit (GRU) variants, are designed to handle sequential data and capture long-term dependencies. They can model temporal dynamics effectively and are suitable for analyzing speech signals. However, RNNs can be computationally expensive and suffer from vanishing or exploding gradients, limiting their ability to capture long-term dependencies effectively.

Despite their advantages, deep learning models also have certain limitations. They require a large amount of labeled data for training, which can be challenging to obtain in the domain of speech emotion. Additionally, deep learning models are prone to overfitting, especially when dealing with limited training data. Furthermore, they often lack interpretability, making it difficult to understand the learned representations and provide explanations for the classification decisions.

\subsection{GCN model}
The Graph Convolution Network (GCN) model is a deep learning architecture that overcomes some of the limitations of traditional deep learning models, such as Convolutional Neural Networks (CNNs) and Bidirectional Long Short-Term Memory (BiLSTM) models.

One challenge that GCN addresses is the lack of consideration for the underlying graph structure in data. Traditional deep learning models like CNNs and BiLSTMs treat the input data as grid-like or sequential, respectively, without explicitly capturing the relationships between data points. In contrast, GCN incorporates graph structures to model relationships and dependencies among the data. By leveraging the graph representation, GCN can capture both local and global dependencies, allowing for a more holistic understanding of the data.

Furthermore, GCN can effectively handle irregular and non-grid data, which is common in many real-world applications. Traditional deep learning models like CNNs rely on fixed grid structures and struggle to handle irregularly structured data. GCN, on the other hand, is designed to handle graph-structured data, making it suitable for various domains where data is inherently non-grid-like, such as social networks, biological networks, and even speech data.

Another advantage of GCN is its ability to capture and propagate information through multiple layers. CNNs are primarily designed for local feature extraction, while BiLSTMs focus on capturing temporal dependencies. However, GCN can capture both local and global dependencies across multiple layers, allowing for a more comprehensive representation of the data. This ability to capture multi-scale dependencies is particularly useful in complex tasks like speech emotion recognition, where emotions may be expressed through various scales of features in speech signals.

Moreover, GCN can handle variable-sized graphs and adaptively update node representations based on their neighborhood information. This adaptive and flexible nature of GCN allows it to effectively model varying relationships between data points, addressing the limitations of traditional models that may struggle with dynamic or changing relationships.

In summary, GCN overcomes the limitations of traditional deep learning models by explicitly incorporating graph structures, handling irregular data, capturing multi-scale dependencies, and adapting to varying relationships. These advantages make GCN a promising approach for tasks like speech emotion recognition, where capturing complex relationships and dependencies among speech features is crucial.

\subsection{HuBERT model}

HuBERT (Hidden Unit BERT) is a deep learning model specifically designed to overcome the limitations of traditional deep learning models in the context of speech recognition tasks. It is pre-trained on a large amount of unlabeled audio data, allowing it to learn powerful representations of speech signals.

One key problem that HuBERT addresses is the need for large amounts of labeled data. Traditional deep learning models often require extensive labeled datasets for training, which can be costly and time-consuming to obtain. In contrast, HuBERT leverages unsupervised pre-training on unlabeled audio data, enabling it to learn general acoustic representations without relying on explicit labels. This allows for more efficient and cost-effective training.

Another challenge HuBERT tackles is the lack of interpretability in deep learning models. Traditional deep learning models, such as CNNs and RNNs, often lack transparency in their internal representations, making it difficult to understand the learned features and provide explanations for classification decisions. HuBERT, being based on the BERT architecture, incorporates the transformer model, which enables better interpretability. The attention mechanism in the transformer allows for understanding the importance and relationships of different parts of the input, providing insights into the learned representations.

Additionally, HuBERT benefits from the language modeling capabilities of BERT. While traditional deep learning models focus solely on acoustic features, HuBERT can also capture contextual information and semantic relationships within speech data. This allows for a more comprehensive understanding of emotions in speech, as linguistic context plays a crucial role in emotional expression.

By leveraging unsupervised pre-training, interpretability, and the ability to capture linguistic context, HuBERT addresses the limitations of traditional deep learning models in speech recognition tasks. It offers improved performance, more efficient training, and a deeper understanding of emotional expressions in speech.

\chapter{Literature Review} 

\label{Literature} 

\lhead{Chapter \ref{Literature}. \emph{Literature Review}} 
 \section{Speech Emotion Recognition using Traditional Machine Learning Approaches}
    Speech Emotion Recognition (SER) using traditional machine learning approaches involves the application of techniques such as Support Vector Machines (SVM), Hidden Markov Models (HMM), and Gaussian Mixture Models (GMM) for emotion classification. These models utilize handcrafted features extracted from speech signals to train classifiers. SVM separates different emotion classes using a hyperplane, while HMM models capture temporal dependencies and transitions between emotional states. GMM represents speech data as a mixture of Gaussian distributions. Although these traditional machine learning approaches have been used successfully in SER, they rely heavily on manually engineered features and may struggle with capturing complex emotional expressions and long-term dependencies in speech.
    \subsection{SER using Support Vector Machine}
    \justify
    SVM is a supervised machine learning algorithm to classify data into different categories. In the paper, \cite{jain2020speech} a method is described by using a set of features, including spectral features (Mel-frequency cepstral coefficients or MFCCs), prosodic features (such as pitch and energy), and linguistic features (such as word count and type-token ratio) as input for the SVM classifier which is trained to predict the emotion of the speech signal. The author also achieved an accuracy of 75\% on 5-class emotion recognition. SVM can actually capture relevant characteristics of speech signals if a set of features is suitable but it can’t be always the case.
    \begin{figure}[!ht]
    \centering
     \includegraphics[width=0.8\textwidth, height=0.5\textwidth]{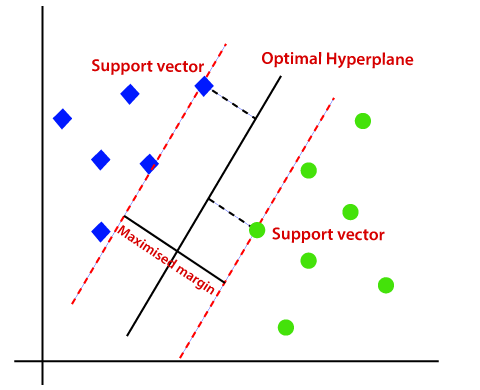}
      \caption {\emph{Proposed method for Speech Emotion Recognition using SVM\cite{jain2020speech}}}
    \end{figure}

    For high-dimensional features, can cause difficult to apply SVM, as high-dimensionality may degrade the performance of the model). Also as SVMs are more pruned for linear decision boundaries, this may limit the ability for non-linear relationships between input features and output labels.
    
    \justify

    \subsection{SER using Hidden Markov Model}
    \justify
     HMM is basically a probabilistic model to determine the sequence in the speech signals and is helpful to represent the sequence of observations of the signals. In the work, \cite{nwe2003speech} author was able to capture total accuracy of 72\% with a 5-class classification task. But limited modeling capacity, as the model's current state, depends on the previous state. Computationally intensive, when the number of states of the model is large, the data is high-dimensional, causing difficulty to apply on real-time applications.

    \subsection{SER using Gaussian Mixture Model}
    \justify
    This work, \cite{Cheng2012/08} is also similar to HMM with improvement in handcrafted acoustic feature analysis and probabilistic classification of emotion but also described the difference between male and female emotional expressions. The below figure explains the idea of Males having more recognition rate than females in 5 Gaussian components.
    \begin{figure}[!hbt]
    \centering
    \includegraphics[scale=1.3]{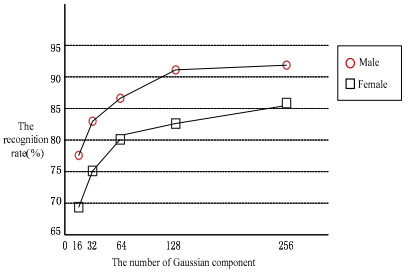}
      \caption {\emph{Accurate identification rate of different genders\cite{Cheng2012/08}}}
    \end{figure}
    The limitation of receiving contextual information yields the need for using more robust and efficient models though.
    
\section{Speech Emotion Recognition using Deep Learning Approaches}
    This review work \cite{jahangir2021deep} shows recent different approaches for SER using deep learning techniques, before coming to deep learning conventional machine learning approaches contained the feature extraction techniques of discriminative and relevant features as they lead to convenient learning of supervised classification algorithms that can more accurately predict emotions. That's why handcrafted feature learning techniques were the trend before the coming of deep learning. Automatic feature learning techniques of DL hugely improved the time constraint of feature learning methods.
     \begin{figure}[!hbt]
    \centering
    \includegraphics[scale=0.7]{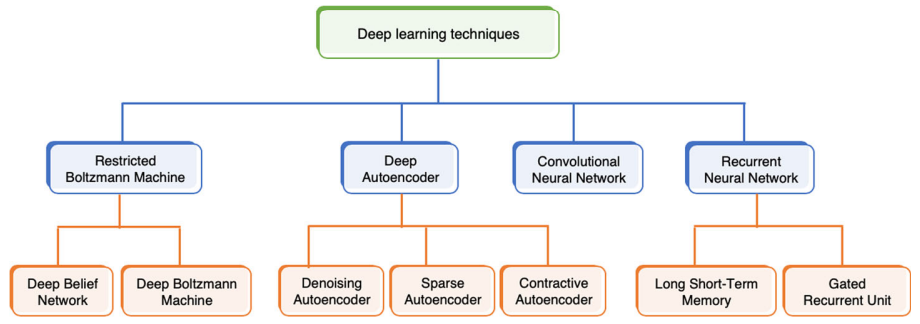}
      \caption {\emph{Different Deep Learning Approaches for SER\cite{jahangir2021deep}}}
    \end{figure}
    \subsection{SER using Generic Deep Neural Network\cite{jahangir2021deep}}
    \justify
    Generic Deep learning approaches basically consist of a bunch of hidden layers, including input layer and output layers to correctly identify weights and features from the network and successfully classify them. Some old traditional DL approaches were using restricted Boltzmann machine architecture (RBM), and Deep Belief Network Architecture (DBN) consists of a bunch of RBMs, The more improved version of these two later came named Deep Boltzmann Machine Architecture. The benefit of DBM was to learning ability was fast and representation was effective due to layer-wise pre-training.
    \begin{figure}[!hbt]
    \centering
    \includegraphics[scale=0.7]
    {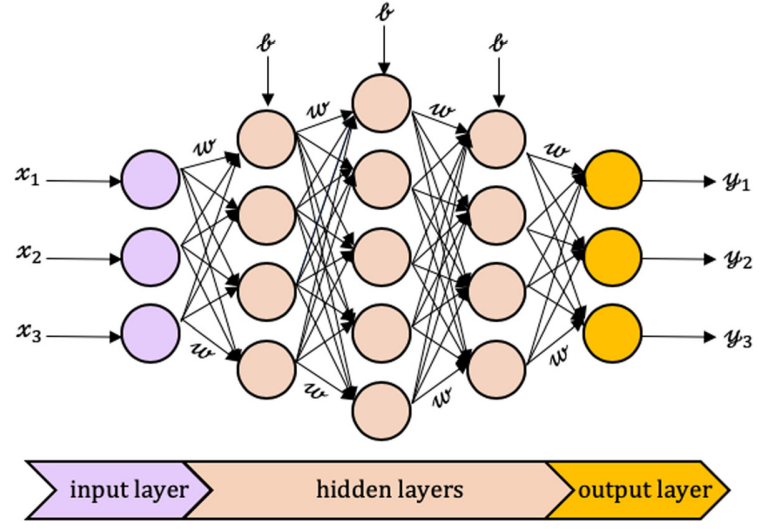}
      \caption {\emph{Generic Deep Neural Network Architecture\cite{jahangir2021deep}}}
    \end{figure}
    \begin{figure}[!hbt]
    \centering
    \includegraphics[scale=0.6]
    {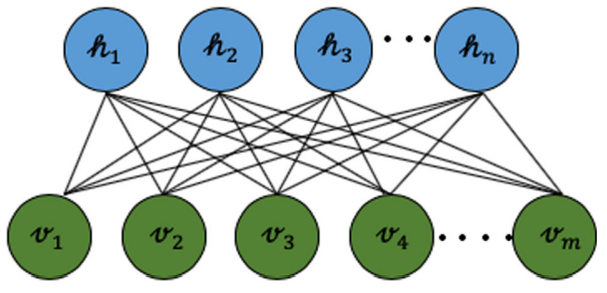}
      \caption {\emph{Restricted Boltzmann Machine Architecture\cite{jahangir2021deep}}}
    \end{figure}
    \begin{figure}[!hbt]
    \centering
    \includegraphics[scale=0.6]
    {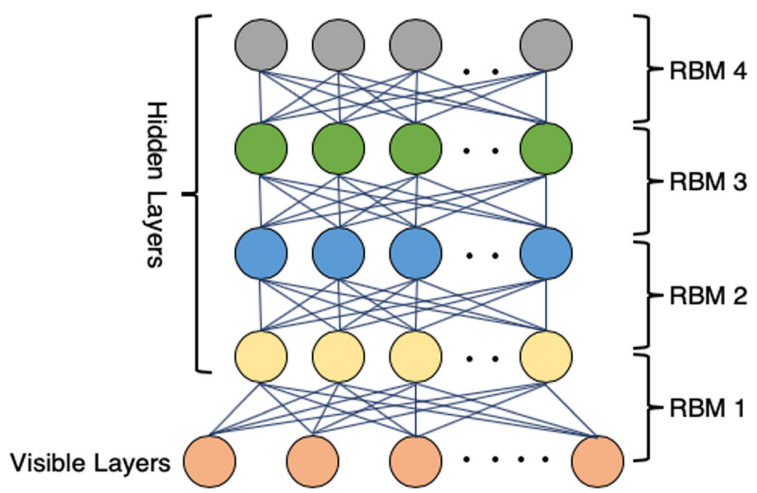}
      \caption {\emph{Deep Belief Network Architecture\cite{jahangir2021deep}}}
    \end{figure}
    \begin{figure}[!hbt]
    \centering
    \includegraphics[scale=0.7]
    {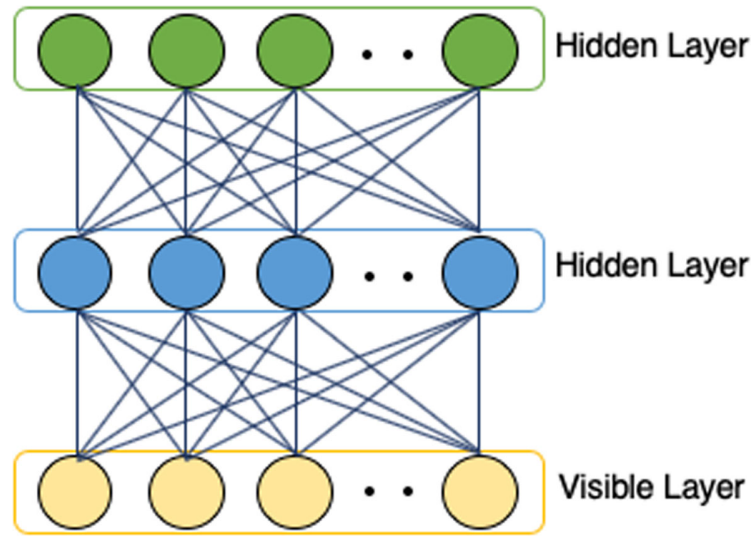}
      \caption {\emph{Deep Boltzmann Machine Architecture\cite{jahangir2021deep}}}
    \end{figure}

    \subsection{SER using Deep autoencoder}
    The autoencoder\cite{bhavan2020deep} technique contributes to having an encoder and decoder in the network, where the encoder is used to capture hidden feature information of input speech data, and the decoder is used to reduce the error percentage of the network.
    \begin{figure}[!hbt]
    \centering
    \includegraphics[scale=0.7]
    {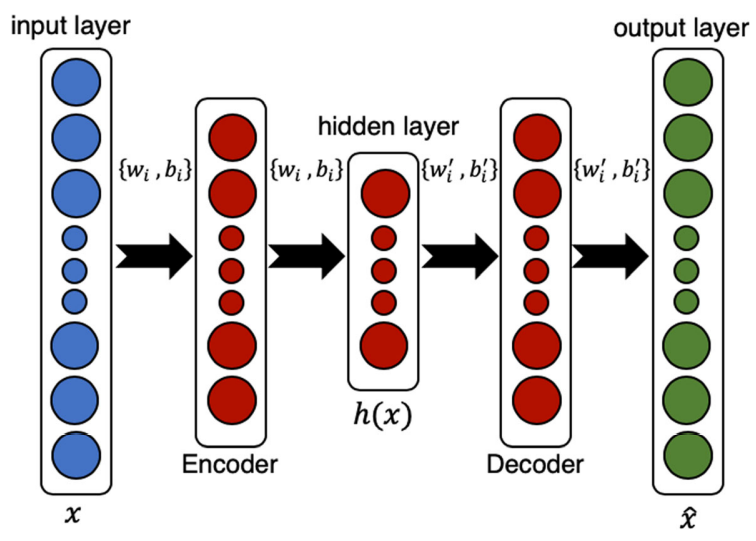}
      \caption {\emph{Deep Autoencoder Architecture\cite{jahangir2021deep}}}
    \end{figure}

    \subsection{SER using Convolutional Neural Network}
    CNN\cite{chauhan2021speech} model general case having three different layers (convolutional, pooling and fully connected layer) successfully capture the class difference of classification problem, where the main layer convolutional comprises the neurons that capture the image spectrograms of speech with aid of a linear filter. 
    \begin{figure}[!hbt]
    \centering
    \includegraphics[scale=0.7]
    {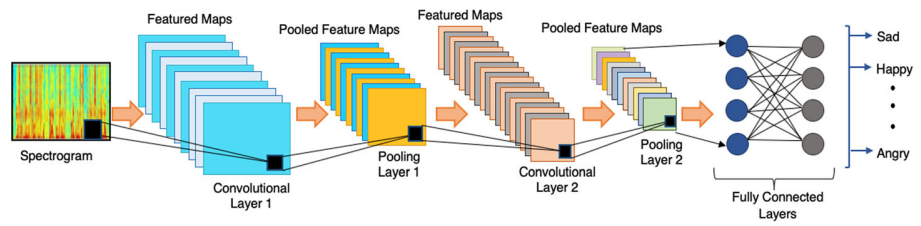}
      \caption {\emph{Convolutional Neural Network Architecture\cite{chauhan2021speech}}}
    \end{figure}

    Then pooling layer does different types of pooling mechanisms including average pooling, stochastic pooling, average pooling, etc. Recently, max pooling showed better performance, hence max pooling technique is now broadly used for the pooling layer due to its structural rigidity of slight variations in speech utterances.

    Then comes the fully connected layer after many convolutional and pooling layers. Fully connected layers connected with classifiers take the master feature vector for classification then successfully classify emotional expressions.

    \subsection{SER using Recurrent Neural Network}
    Speech can be represented as a 2-D graph of frequency and time. That's why speech databases are also known as time series data. Sequential information is important hence. Temporal features thus can be captured by sequential representation understanding models like RNN\cite{li2021speech}. Different variations of RNN are now used frequently like Long Short Term Memory (LSTM) for capturing long-term temporal dependencies of speech. Though LSTM proves to provide better performance for speech, it eventually contributes to the complexity and exploding gradient problems. 
    \begin{figure}[!hbt]
    \centering
    \includegraphics[scale=0.8]
    {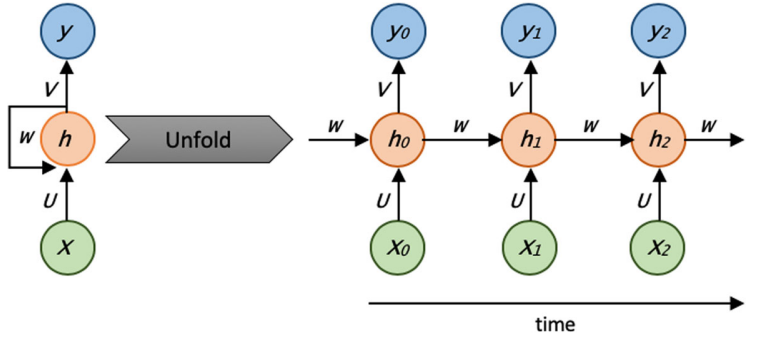}
      \caption {\emph{Recurrent Neural Network Architecture\cite{han2021speech}}}
    \end{figure}

    \subsection{SER using Evolution of Deep Learning Techniques}
    Over the years, DL techniques\cite{issa2020speech} for SER improved hugely and it can be broadly categorized into three, Generative, Discriminative, and Hybrid. In recent times due to the complexity of Discriminative models, Hybrid models thus perform more effectively than others. Taxonomy is given below for the evolution of models of SER.
    \begin{figure}[!hbt]
    \centering
    \includegraphics[scale=1]
    {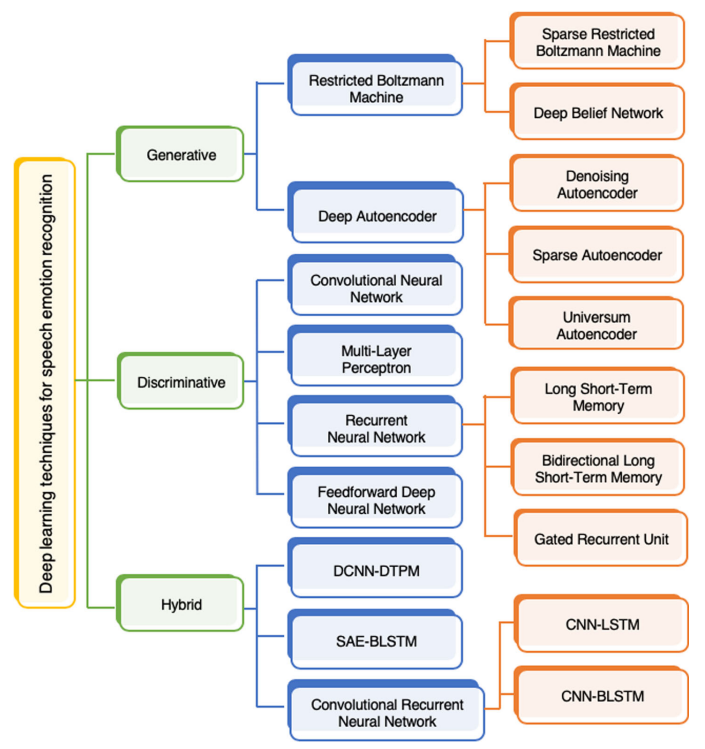}
      \caption {\emph{Taxonomy of DL techniques for SER\cite{issa2020speech}}}
    \end{figure}
    
    \subsection{SER using Different Benchmark Literatures of recent times}
    \justify
    This paper \cite{amiriparian2021impact} explores the impact of word error rate (WER) on the performance of acoustic-linguistic speech emotion recognition models. The authors analyze the performance of deep learning models for speech emotion recognition on a dataset with different levels of WER and compare their results to those obtained using traditional machine learning approaches. They find that the performance of deep learning models is more robust to WER than that of traditional machine learning approaches and that these models are able to achieve good performance even when the WER is high. The authors also explore the impact of different types of errors on the performance of the models and find that recognition errors that affect the content of the speech signal has a larger impact on the performance of the models than errors that do not affect the content.
    \begin{figure}[!hbt]
    \centering
    \includegraphics[scale=0.8]{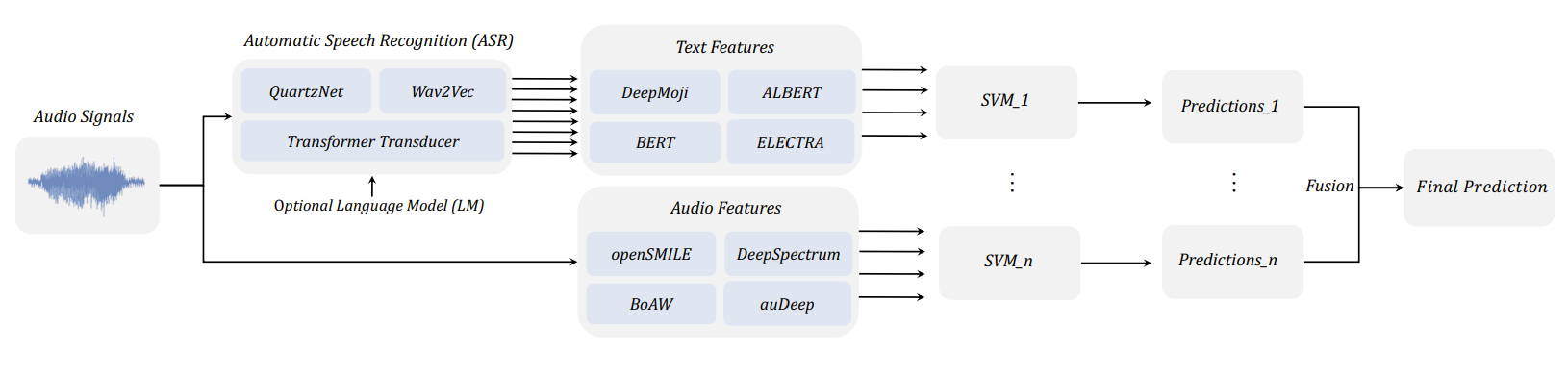}
      \caption {\emph{A general overview of our emotion recognition by multi-modal fusion of different trained Support Vector Machines (SVMs).\cite{amiriparian2021impact}}}
    \end{figure}
    Overall, this paper suggests that deep learning models are more robust to WER than traditional machine learning approaches for speech emotion recognition, and highlights the importance of considering the impact of WER on the performance of these models. It also suggests that errors that affect the content of the speech signal may have a larger impact on the performance of these models than other types of errors. The features act as the inputs for Support Vector Machines, which form the basis for our SER analysis (SVMs). Since IEMOCAP is established and suitable for comparison with state-of-the-art approaches, it is the main reason we chose it. Additionally, IEMOCAP contains transcriptions that make it easier to evaluate how the ASR’s WER impacted the final emotion classification.

    \justify
    This paper \cite{scheidwasser2022serab} describes the creation of a multi-lingual benchmark dataset for speech emotion recognition (SER). The authors describe the process of collecting and annotating a large dataset of speech samples in five different languages (English, Spanish, French, German, and Italian) with a total of 7,659 utterances. The dataset includes a range of emotional states, including neutral, happy, sad, angry, and fearful, and was annotated using multiple annotators to ensure high-quality labels. The authors also describe the performance of several state-of-the-art SER models on the dataset and compare their results to those obtained on other datasets. They find that the models achieve good performance on the SERAB dataset and that the performance is similar across the different languages. Overall, this paper presents a new benchmark dataset for SER that includes a wide range of emotional states and languages and provides a useful resource for researchers working in this field. It also demonstrates the effectiveness of state-of-the-art SER models on this dataset and highlights the potential for cross-lingual generalization of these models.
    \begin{figure}[!hbt]
    \centering
    \includegraphics[scale=1.7]{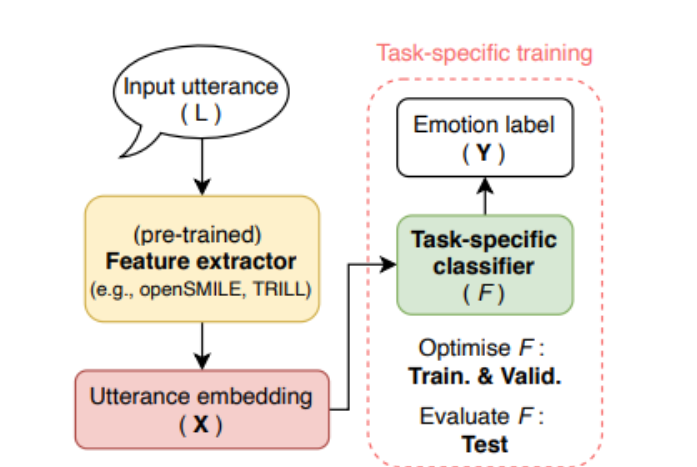}
      \caption {\emph{SERAB evaluation pipeline. The (pre-trained/non-trainable)feature extractor is used to obtain utterance-level embeddings (X) from the input. X are used as input to the task-specific classifier F optimized for predicting the emotion Y expressed in the input.\cite{scheidwasser2022serab}}}
    \end{figure}

    \justify
    This paper \cite{wang2021learning}  presents a new approach for speech emotion recognition (SER) using a multimodal transformer model. The authors propose a multimodal transformer model that takes both speech and text as input and learns to extract features from both modalities. The model includes a mutual correlation module that captures the correlation between the speech and text modalities and allows the model to learn from both modalities simultaneously. The authors evaluate the performance of the proposed model on a speech emotion recognition dataset and compare it to several baseline models. They find that the multimodal transformer model outperforms the baselines and achieves good performance on the task. Overall, this paper presents a new approach for SER that takes advantage of both speech and text modalities and is able to capture the mutual correlation between the two. It demonstrates the effectiveness of the proposed approach and suggests that it could be a useful tool for SER in practical settings.
    \begin{figure}[!hbt]
    \centering
    \includegraphics[scale=1.7]{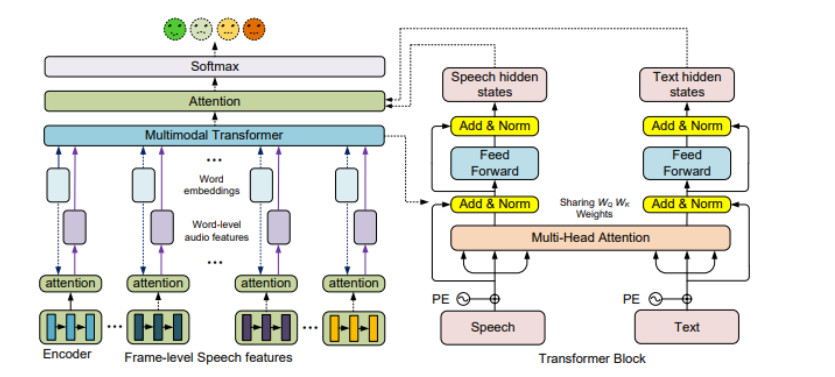}
      \caption {\emph{The architecture of their proposed model. The right part illustrates the multimodal transformer fusion with weights sharing; the left part illustrates the word-level speech representation and the attention-based interaction mechanism.\cite{wang2021learning}}}
    \end{figure}

    \justify
    This paper \cite{ho2020multimodal} presents a new multimodal approach for speech emotion recognition (SER) using a multi-level multi-head fusion attention-based recurrent neural network (MLMF-RNN). The authors propose an MLMF-RNN model that takes both speech and text as input and uses a multilevel attention mechanism to learn from both modalities simultaneously. The model includes multiple attention heads at different levels of the network, which allow it to focus on different aspects of the input and capture more fine-grained information. The authors evaluate the performance of the proposed model on a speech emotion recognition dataset and compare it to several baseline models. They find that the MLMF-RNN model outperforms the baselines and achieves good performance on the task. Overall, this paper presents a new multimodal approach for SER that takes advantage of both speech and text modalities and uses a multi-level attention mechanism to learn from the two modalities simultaneously. It demonstrates the effectiveness of the proposed approach and suggests that it could be a useful tool for SER in practical settings.
    \begin{figure}[!hbt]
    \centering
    \includegraphics[scale=0.9]{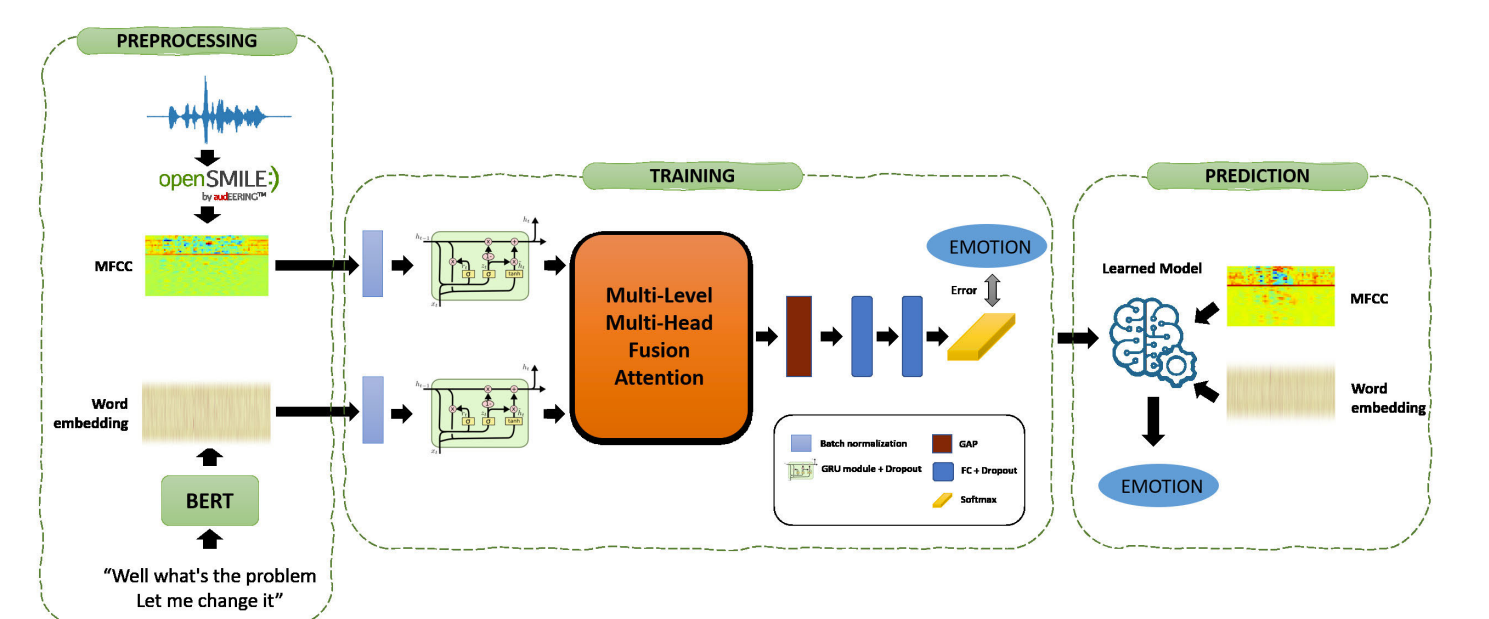}
      \caption {\emph{Overall architecture of the proposed MMFA-RNN model based SER.\cite{ho2020multimodal}}}
    \end{figure}

    \justify
    This paper \cite{heusser2019bimodal} presents a new approach for speech emotion recognition (SER) using pre-trained language models. The authors propose a bimodal SER model that takes both speech and text as input and uses pre-trained language models to extract features from each modality. The model includes a multimodal fusion layer that combines the features from the two modalities and feeds them into a classification layer to predict the emotional state of the speech signal. The authors evaluate the performance of the proposed model on a speech emotion recognition dataset and compare it to several baseline models. They find that the bimodal SER model outperforms the baselines and achieves good performance on the task. However, this paper provides a new approach for SER that takes advantage of both speech and text modalities and uses pre-trained language models to extract features from the two modalities.

    \justify
    This paper \cite{wang2021novel} presents a new end-to-end approach for speech emotion recognition (SER) using stacked transformer layers. The authors propose an SER model that consists of a series of stacked transformer layers that process the raw speech signal and learn to extract features from it. The model includes a self-attention mechanism that allows it to capture long-range dependencies in the speech signal and a multi-head attention mechanism that allows it to focus on different aspects of the input simultaneously. The authors evaluate the performance of the proposed model on a speech emotion recognition dataset and compare it to several baseline models. They find that the stacked transformer SER model outperforms the baselines and achieves good performance on the task. Moreover, this paper presents a new end-to-end approach for SER that uses stacked transformer layers to process the raw speech signal and learn to extract features from it.
    \begin{figure}[!hbt]
    \centering
    \includegraphics[scale=1]{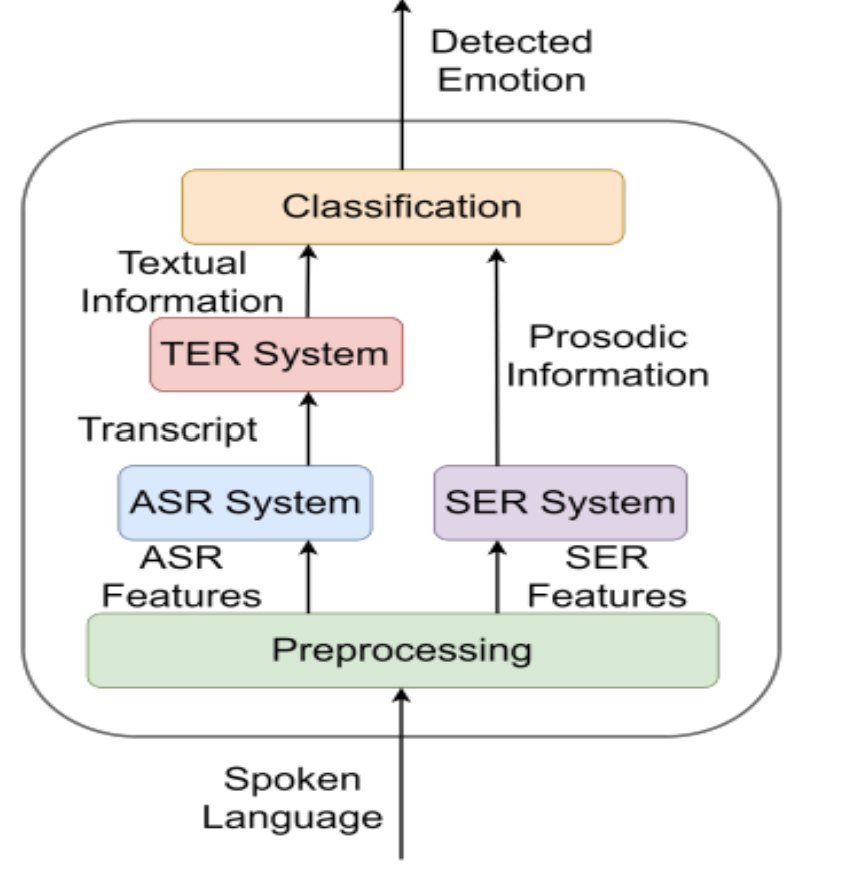}
      \caption {\emph{The IEmoNet framework where each intermediate system (ASR, SER, TER) can be chosen and optimized individually.\cite{wang2021novel}}}
    \end{figure}

    \begin{figure}[!hbt]
    \centering
    \includegraphics[scale=1.7]{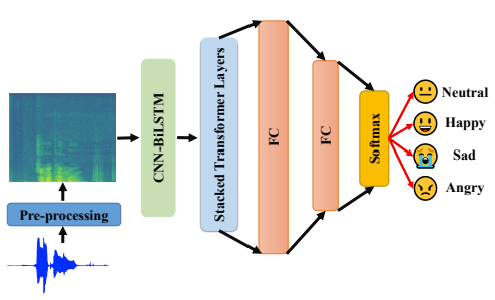}
      \caption {\emph{Our proposed SER architecture.\cite{wang2021novel}}}
    \end{figure}

    \begin{figure}[!hbt]
    \centering
    \includegraphics[scale=2]{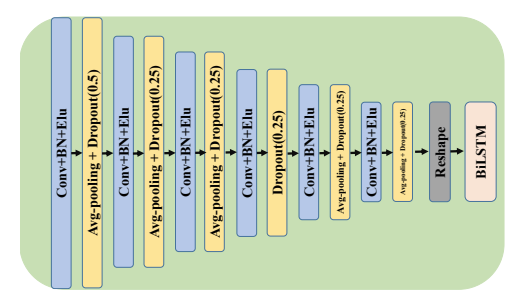}
      \caption {\emph{Detailed architecture of CNN-BiLSTM module in the previous figure\cite{wang2021novel}}}
    \end{figure}

    \justify
    This paper \cite{han2021speech} presents a new approach for speech emotion recognition (SER) using a parallel neural network that combines a residual convolutional neural network (ResNet-CNN) with a transformer network. In this paper, the authors propose a parallel SER model that takes the raw speech signal as input and processes it using both a ResNet-CNN and a transformer network. The output of the two networks is then combined using a fusion layer and fed into a classification layer to predict the emotional state of the speech signal. The authors evaluate the performance of the proposed model on a speech emotion recognition dataset and compare it to several baseline models. They find that the parallel SER model outperforms the baselines and achieves good performance on the task. Overall, this paper presents a new approach for SER that combines a ResNet-CNN and a transformer network to process the raw speech signal and learn to extract features from it. It demonstrates the effectiveness of the proposed approach and suggests that it could be a useful tool for SER in practical settings.
    \begin{figure}[!hbt]
    \centering
    \includegraphics[scale=1.3]{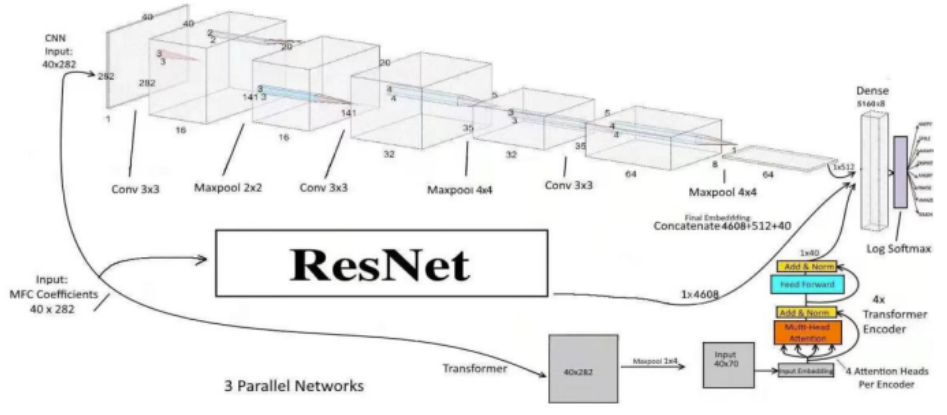}
      \caption {\emph{Architecture of ResNet-CNN model.\cite{han2021speech}}}
    \end{figure}

    \justify
    In this paper, \cite{zhao2019speech}\cite{huang2014speech} the authors propose two SER models that use different types of CNNs (1D and 2D) to extract features from the raw speech signal and an LSTM network to process the features over time. The models are trained end-to-end to predict the emotional state of the speech signal. The authors evaluate the performance of the proposed models on a speech emotion recognition dataset and compare them to several baseline models. They find that the deep CNN LSTM SER models outperform the baselines and achieve good performance on the task. This paper provides two new approaches for SER that use deep CNN LSTM networks to process the raw speech signal and learn to extract features from it. It encapsulates the effectiveness of the suggested approaches and offers that they could be useful tools for SER in practical settings.

    \justify
    This paper \cite{shirian2021compact} proposed a new approach for speech emotion recognition (SER) using a compact graph architecture. In this paper, the authors propose an SER model that takes the raw speech signal as input and processes it using a series of graph convolutional layers. The model includes a self-attention mechanism that allows it to capture long-range dependencies in the speech signal and a fusion layer that combines the features from the graph convolutional layers and feeds them into a classification layer to predict the emotional state of the speech signal. The authors evaluate the performance of the proposed model on a speech emotion recognition dataset and compare it to several baseline models. They find that the compact graph SER model outperforms the baselines and achieves good performance on the task. Overall, this paper presents a new approach for SER that uses a compact graph architecture to process the raw speech signal and learn to extract features from it. It demonstrates the effectiveness of the proposed approach and suggests that it could be a useful tool for SER in practical settings.
    \begin{figure}[!hbt]
    \centering
    \includegraphics[scale=0.7]{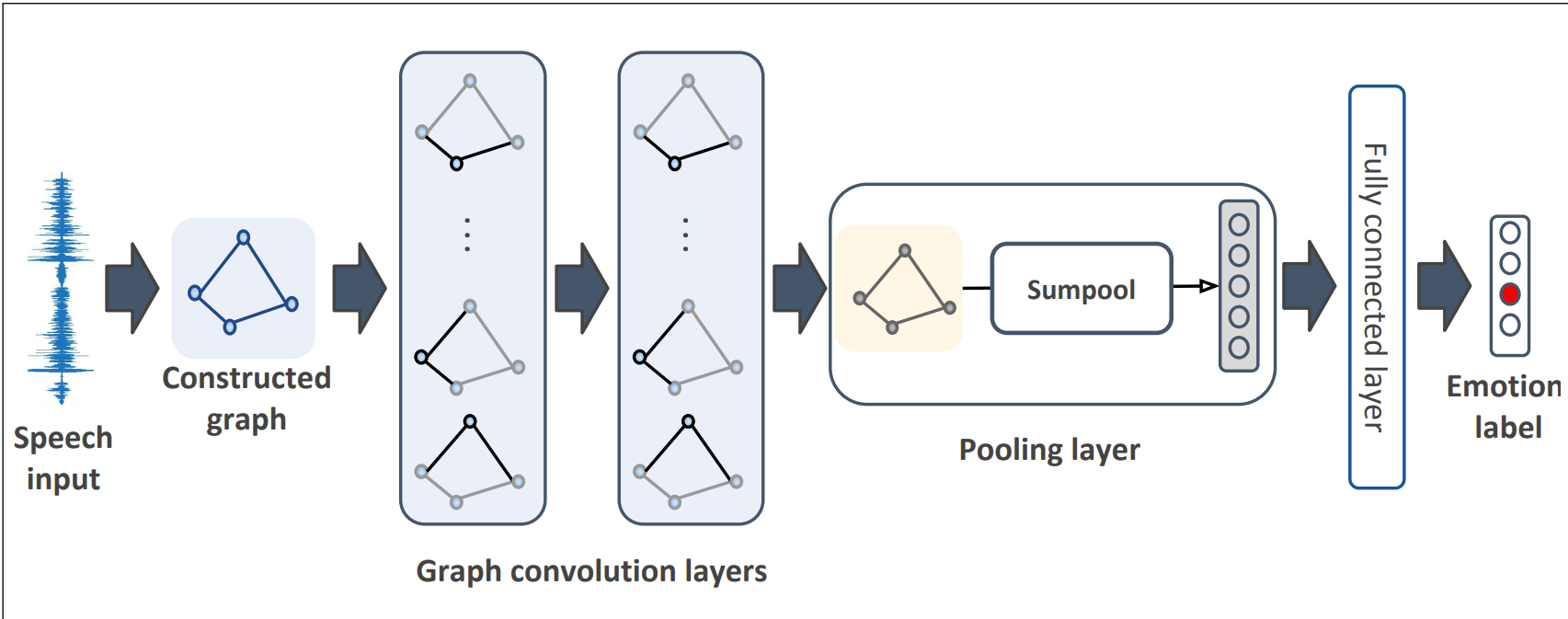}
      \caption {\emph{Proposed graph-based architecture for SER consists of two graph convolution layers and a pooling layer to learn graph embedding from node embeddings to facilitate emotion classification.\cite{shirian2021compact}}}
    \end{figure}

    \begin{figure}[!hbt]
    \centering
    \includegraphics[scale=1]{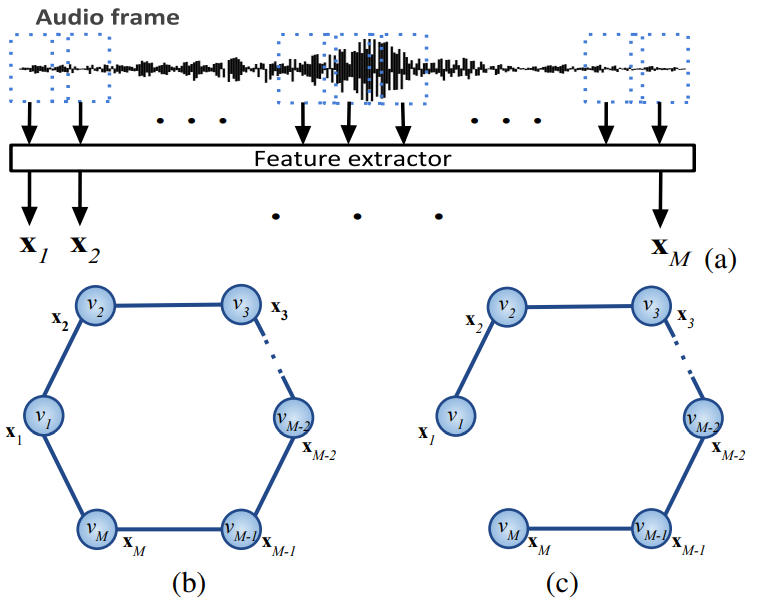}
      \caption {\emph{Graph construction from the speech input. (a) LLDs are extracted as node features xi from raw speech segments; (b) cycle graph, and (c) chain graph.\cite{shirian2021compact}}}
    \end{figure}

    \justify
    The paper ”Multimodal Speech Emotion Recognition and Classification Using Convolutional Neural Network Techniques”\cite{christy2020multimodal} presents a method for recognizing emotions in speech using a multimodal approach, which combines information from multiple modalities such as audio, text, and facial expression.
    The proposed method consists of three main components: a feature extraction module, a feature fusion module, and a classification module. The feature extraction module extracts a set of audio, text, and facial expression features from the input data. The feature fusion module combines the extracted features using a convolutional neural network (CNN) to learn a unified representation of the data. The classification module uses another CNN to predict the emotion label of the input data based on the fused representation. The proposed method was evaluated on a publicly available dataset of speech signals annotated with emotional labels. The results showed that the proposed method outperformed several state-of-the-art methods for speech emotion recognition. The results also showed that the multimodal approach improved the performance of the method compared to using only a single modality. The authors evaluate the performance of the proposed model on a speech emotion recognition dataset and compare it to several baseline models. They find that the multimodal CNN SER model outperforms the baselines and achieves good performance on the task.

\chapter{Proposed Approach} 

\label{methodology} 

\lhead{Chapter \ref{methodology}. \emph{Proposed Approach}} 
The proposed methodology for our research involves the following steps. Firstly, we take the speech signal and create two separate modules: one for the spectral representation of the audio and another for the textual representation of the audio data. The textual module is processed using the Graph Convolution Network (GCN) model, which captures relationships and dependencies among the textual features. The spectral module is processed using the HuBERT model, a pre-trained model specifically designed for speech recognition tasks. After obtaining individual classification scores from each module, we perform score-level fusion to combine the scores. Finally, we classify the speech emotion based on the fused scores, providing a comprehensive and integrated approach to speech emotion recognition.
\begin{figure}[H]
    \centering
    \includegraphics[scale=0.115]
    {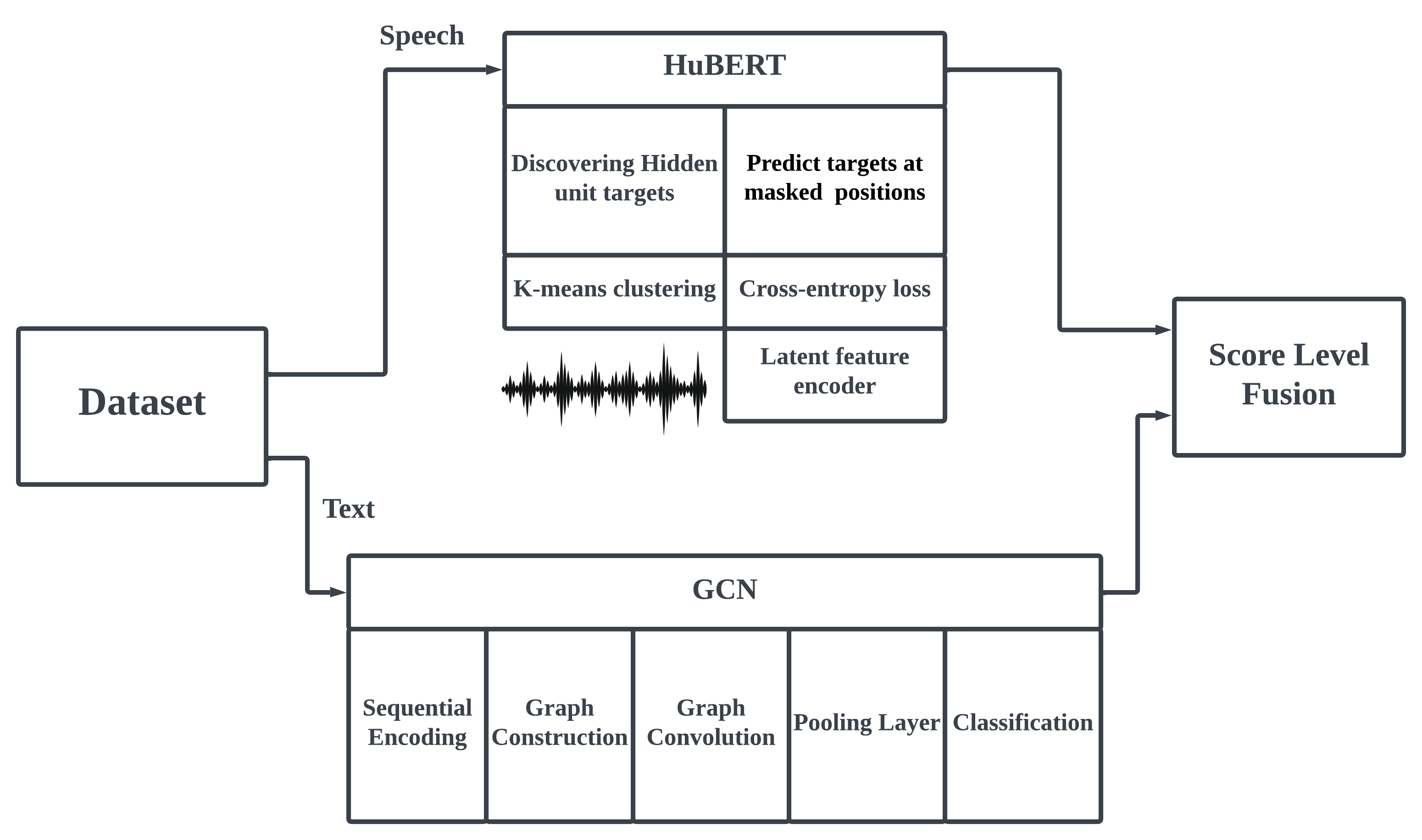}
      \caption {\emph{Our Proposed Architecture}}
\end{figure}
\section{Dataset Preparation}
    We mainly used two popular datasets that are frequently used for emotion recognition tasks, IEMOCAP and RAVDESS. The below section describes more about these datasets.
    \subsection{IEMOCAP}
    The IEMOCAP\cite{busso2008iemocap} dataset is a popular choice for emotion recognition because it contains valuable conversational emotion information. It has a wide range of audio and video recordings where actors engage in scripted and spontaneous conversations, showing genuine emotions. This dataset is unique because it captures the subtle details of emotions during natural conversations. Researchers and practitioners find it useful for developing models that can accurately understand and classify emotions in real-life conversations. Its focus on capturing conversational dynamics makes it a valuable resource for advancing the field of emotion recognition.
    IEMOCAP dataset descriptions are:
    \begin{table}[H]
    \caption{IEMOCAP Dataset Description}
    \begin{tabular}{|l|l|l|l|}

    \hline
    \textbf{Column} & \textbf{Type} & \textbf{Values}                                                                                                  & \textbf{Descriptions}                                                                                             \\ \hline
    'session'       & int           & 1, 2, 3, 4, 5                                                                                                    & \begin{tabular}[c]{@{}l@{}}dialogue sessions in the\\ database\end{tabular}                                       \\ \hline
    'method'        & str           & 'script', 'impro'                                                                                                & \begin{tabular}[c]{@{}l@{}}the method of emotion \\ elicitation\end{tabular}                                      \\ \hline
    'gender'        & str           & 'M', 'F'                                                                                                         & gender of the speaker                                                                                             \\ \hline
    'emotion'       & str           & \begin{tabular}[c]{@{}l@{}}'neu', 'fru', 'sad', \\ 'sur', 'ang', 'hap',\\ 'exc', 'fea', 'dis','oth'\end{tabular} & annotated emotion                                                                                                 \\ \hline
    'n\_annotators' & int           & -                                                                                                                & number of annotators                                                                                              \\ \hline
    'agreement'     & int           & 2, 3, 4                                                                                                          & \begin{tabular}[c]{@{}l@{}}number of annotators who\\ agree to this label\end{tabular}                            \\ \hline
    'path'          & str           & 'path/to/file/'                                                                                                  & \begin{tabular}[c]{@{}l@{}}path to the .wav file, relative to \\ "IEMOCAP\_full\_release/" directory\end{tabular} \\ \hline
    \end{tabular}
    \end{table}

   \subsection{RAVDESS}
   \justify
   Due to its applicability and thoroughness, the RAVDESS\cite{livingstone2018ryerson} dataset is an appealing option for emotion identification. It consists of a variety of audiovisual recordings that feature experienced performers and allow for the recording of a spectrum of emotional emotions. The dataset offers a precise and well-defined emotional ground truth by concentrating on certain emotional categories including happiness, sadness, rage, and neutral moods. It ensures the inclusion of a variety of demographic characteristics thanks to the equal representation of male and female performers as well as differences in age and cultural background. Researchers and practitioners may examine the synergistic impact of multimodal signals on emotion identification algorithms since both auditory and visual modalities are readily available. In addition, its open-access features and established benchmarks make it a useful tool for assessing and contrasting various approaches in the field of emotion recognition.

   \justify
   RAVDESS dataset descriptions are:
   Filename extensions are 03-01-06-01-02-01-12.wav representations for,
    Filename identifiers

    \begin{itemize}
        \item Modality (01 = full-AV, 02 = video-only, 03 = audio-only).
        \item Vocal channel (01 = speech, 02 = song).
        \item Emotion (01 = neutral, 02 = calm, 03 = happy, 04 = sad, 05 = angry, 06 = fearful, 07 = disgust, 08 = surprised).
        \item Emotional intensity (01 = normal, 02 = strong). NOTE: There is no strong intensity for the 'neutral' emotion.
        \item Statement (01 = "Kids are talking by the door", 02 = "Dogs are sitting by the door").
        \item Repetition (01 = 1st repetition, 02 = 2nd repetition).
        \item Actor (01 to 24. Odd numbered actors are male, even-numbered actors are female).
    \end{itemize}

    \justify
    Preprocessing for the audio same goes likewise IEMOCAP. Basically, the pipeline is Audio preprocessing, Graph Construction, Node feature representation and graph convolution.

    \justify
    Extract relevant audio features from the raw audio files in the RAVDESS dataset. Commonly used features for speech include Mel-frequency cepstral coefficients (MFCCs), pitch, energy, and spectrograms. These features capture the acoustic characteristics of the speech signals. These features are preprocessed by the internal encoder of HuBERT.
\section{GCN Module}
\begin{itemize}
    \item \textbf{Building the graph:}The graph building process involves the following steps:
    \begin{enumerate}
        \item \textbf{Preprocessing the text:}Tokenize the sentence into individual words and apply any necessary text cleaning techniques such as removing punctuation, empty spaces, and converting words to lowercase.
        \item \textbf{Creating word vectors:} Use a pre-trained GloVe word embedding model to represent each word as a vector. This creates a numerical representation of the meaning or context of each word.
        \item \textbf{Graph Construction:} Each word vector becomes a node in the graph. Connect the nodes (words) with edges based on their relationships within the sentence. The relationships can be established through various techniques such as dependency parsing or syntactic analysis. For example, you can connect nodes that are adjacent to each other in the sentence, representing the sequential order of the words.
        \begin{itemize}
            \item \textbf{Nodes:} Each word vector becomes a node in the graph. We have nodes for each word in the sentence.
            \item \textbf{Edges:} Connect the nodes based on their relationships within the sentence. Let's assume we connect nodes that are adjacent to each other in the sentence.
            \item \textbf{Assigning weights to edges:} We calculate the cosine similarity between word vectors to determine edge weights based on proximity. We use values between 0 and 1, where a higher value indicates a stronger relationship.
        \end{itemize}
        \item \textbf{Graph representation:} The graph can be represented using an adjacency list or an adjacency matrix. The graph represents the relationships between individual words in the sentence, and the emotion labels assigned to the sentence can guide the analysis and interpretation of the graph to recognize the emotions expressed.
    \end{enumerate} 
    \item \textbf{Graph convolution:}The graph convolution operation and the pooling layer in graph convolutional networks work as below:
    \begin{enumerate}
        \item \textbf{Graph Convolution Operation:}The graph convolution operation is defined using the formula: 
        \begin{equation}
            H_{hat} = (U_{transpose} * X) * (U_{transpose} * W)
        \end{equation}
        , where $H_{hat}$ represents the transformed node features.H is computed as 
        \begin{equation}
            H = U * H_{hat}
        \end{equation}, where U is the graph Fourier transform matrix.The propagation of graph convolution at the k-th layer is given by: 
        \begin{equation}
            H^{k+1} = U * (U_{transpose} * H^{k}) * (U_{transpose} * W^{k})
        \end{equation}. This formulation leverages the graph Fourier transform (GFT) and avoids computationally expensive eigen decomposition for arbitrary graph structures. The learnable parameters are the weights W in the convolution operation.
        \item \textbf{Pooling Layer:}The goal of the pooling layer is to obtain a graph-level representation  from the node-level embeddings at the final layer. Common choices for graph pooling functions are mean, max, and sum pooling. Max and mean pooling may not preserve the underlying graph structure information effectively, while sum pooling has been shown to be a better alternative. The pooling operation aggregates the node-level embeddings to obtain a graph-level representation that can be further used for classification or downstream tasks. Overall, the graph convolution operation applies the learned convolution kernel to update node features based on their relationships in the graph, and the pooling layer aggregates node-level embeddings to obtain a graph-level representation for classification or other tasks.\pagebreak
        \item \textbf{Classification:}The pooling layer is followed by one fully-connected layer which produces the classification labels. Our GCN model is trained with cross-entropy loss.
    \end{enumerate}
\end{itemize}

\begin{figure}[H]
    \centering
    \includegraphics[scale=0.1]
    {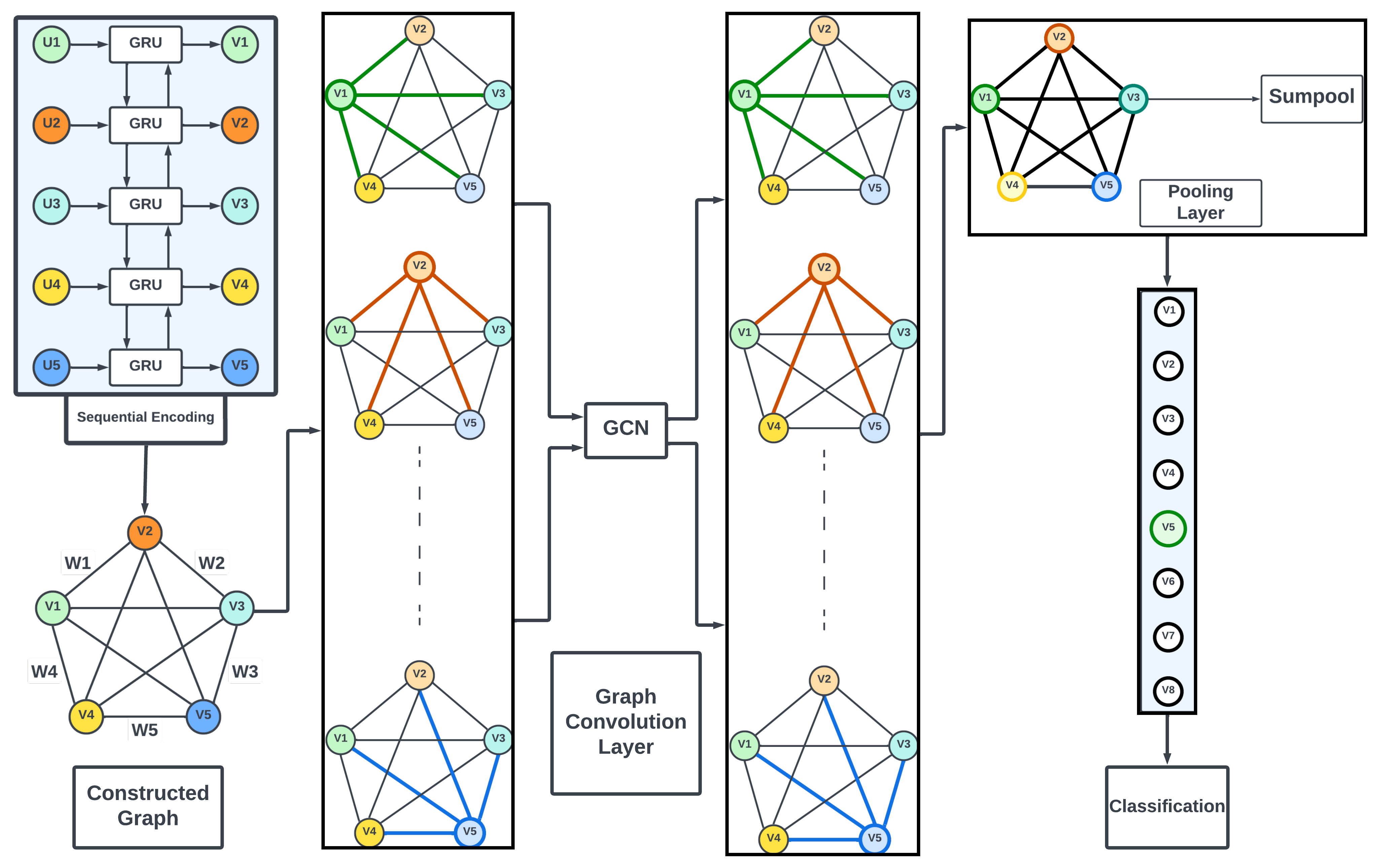}
      \caption {\emph{Our proposed GCN architecture}}
\end{figure}

\section{HuBERT Module}
The fundamental principle is to convert speech data into a more "language-like" structure by identifying discrete hidden units (the Hu in the name). The words or tokens in a written sentence could be compared to these hidden units. We may use the same potent models, such BERT, that are available for natural language processing by representing speech as a series of discrete units.

The technique is based on the DeepCluster paper in computer vision, where images are classified into a predetermined number of clusters before being used again as "pseudo-labels" for self-supervised model training. HuBERT uses short audio segments (25 milliseconds) instead of images for its clustering, and the generated clusters serve as the hidden units that the model will be taught to predict.

\textbf{Training process}
The two parts of the training process alternate between creating pseudo-targets by clustering and trying to forecast them at hidden locations through prediction.

\begin{itemize}
    \item \textbf{Step 1: Discover “hidden units” targets through clustering}\\The concealed units (pseudo-targets) must first be extracted from the audio's raw waveform. Each 25-millisecond audio fragment is divided into K clusters using the K-means algorithm. All audio frames associated to each recognized cluster will subsequently be given the unit label for the hidden unit that was created for it. Then, each hidden unit is mapped to its appropriate embedding vector, which can be utilized to produce predictions in the subsequent phase. 
 
    The most crucial clustering decision is which characteristics to turn the waveform into. Mel-Frequency Cepstral Coefficients (MFCCs), which have been demonstrated to be somewhat effective for speech processing, are employed for the first clustering phase. However, representations from a middle layer of the HuBERT transformer encoder (from the prior iteration) are reused for further clustering phases.

   The HuBERT BASE model is only trained for two iterations, and during the second iteration, clustering is performed using the sixth transformer layer. The ninth transformer layer from the second iteration of the BASE model is also used to train HuBERT LARGE and X-LARGE for a third iteration.

    To capture targets with variable granularity (for example, vowel/consonant vs. sub-phone states), we experiment with combining clustering of various sizes as well as multiple clustering with various numbers of clusters. They demonstrate that employing cluster ensembles can marginally enhance performance. The original publication contains additional information on this.

    \item \textbf{Step 2: Predict noisy targets from the context}\\The second step involves training with the masked language modeling aim, just like with the original BERT. The model is requested to forecast the objectives at positions where about 50\% of the transformer's encoder input features are hidden. Prediction logits are calculated for this by computing the cosine similarity between the transformer outputs (projected to a lower dimension) and each hidden unit embedding from all feasible hidden units. Then, incorrect predictions are punished using cross-entropy loss. Since it has been demonstrated to perform better when utilizing noisy labels, the loss is only applied to the spots that are masked. By attempting to forecast only masked targets, only unmasked targets, or both at once, the authors experimentally demonstrate this.

\end{itemize}

\begin{figure}[H]
    \centering
    \includegraphics[scale=0.225]
    {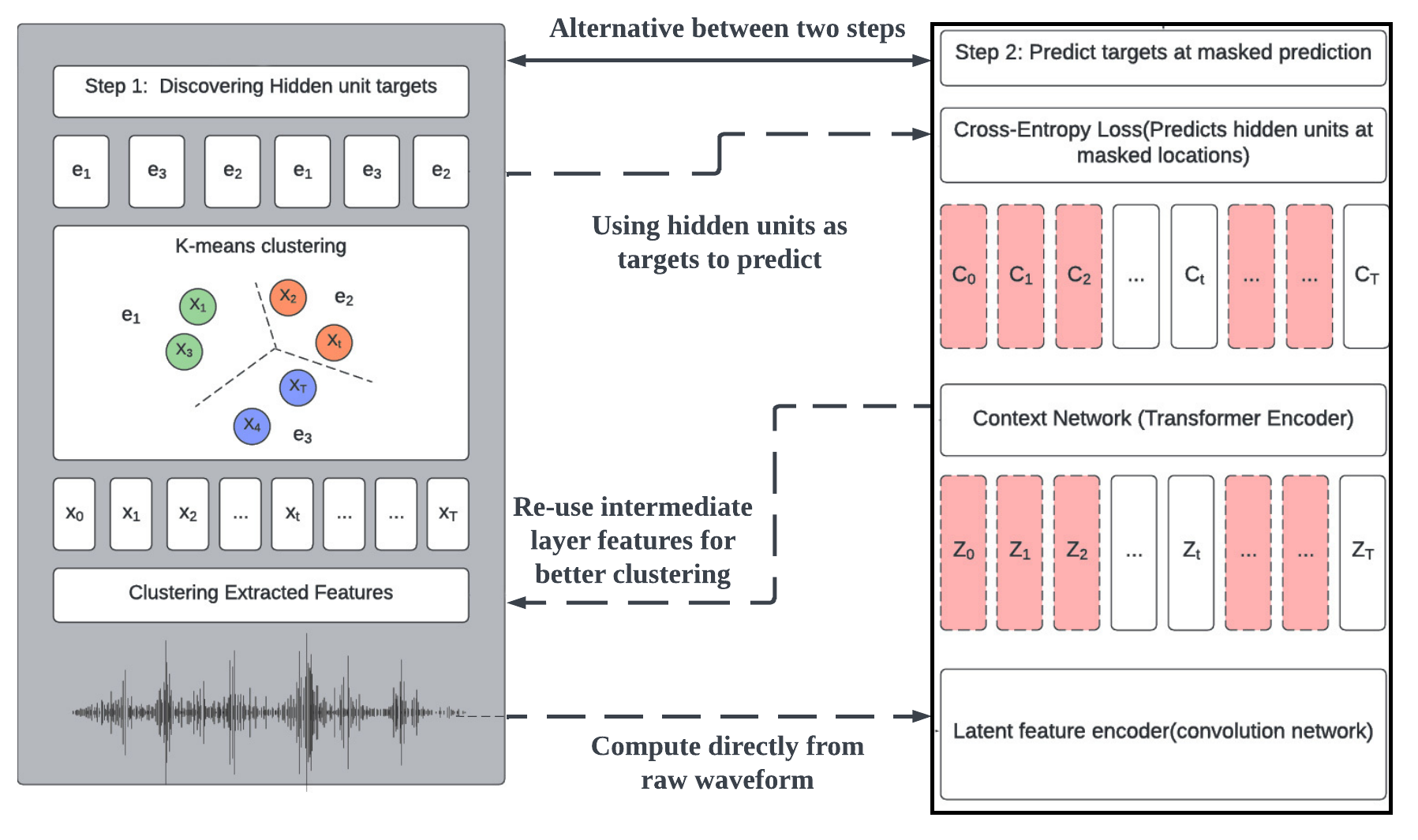}
      \caption {\emph{Our Proposed HuBERT Architecture}}
\end{figure}

\section{Score level Fusion}
In score-level fusion, the predicted scores from different modalities (in this case, GCN for textual representation and HuBERT for spectral representation) are combined to make a final prediction. The softmax function is commonly used in score-level fusion to convert the raw scores into probability distributions over the possible classes.
Here's the step-by-step process of using softmax for score-level fusion:
\begin{enumerate}
    \item Obtain the predicted scores from the GCN and HuBERT models for each emotion class.\\Example: \\GCN predictions: [0.3, 0.6, 0.1, 0.2] (example scores for each class)\\HuBERT predictions: [0.2, 0.4, 0.6, 0.8] (example scores for each class)
    \item Apply the softmax function to each set of scores separately.\\GCN softmax: [0.214, 0.380, 0.131, 0.275] (example probabilities for each class after applying softmax)\\HuBERT softmax: [0.118, 0.193, 0.288, 0.401] (example probabilities for each class after applying softmax)
    \item Compare the probabilities for each class between the two modalities and select the class with the highest probability.
    \\Combined probabilities: [0.214, 0.380, 0.288, 0.401] (taking the maximum probability for each class)
\end{enumerate}

Finally, we make the final prediction based on the class with the highest probability.

In this case, the class with the highest probability is chosen as the final predicted emotion class. Using the softmax function in score-level fusion ensures that the predicted scores are transformed into meaningful probabilities, allowing for easier interpretation, decision-making, and model calibration.

\chapter{Experimental Design} 

\label{experiment} 

\lhead{Chapter \ref{experiment}. \emph{Experimental Design}} 

\justify
This chapter discusses all the experiments we did conduct throughout our research, like dataset selection, preprocessing, model architecture, training and evaluation, cross-validation and hyperparameter tuning, and comprehensive analysis. We designed to experiment on both the HuBERT and GCN models as HuBERT  is a very powerful model to capture the spectral information designed for speech and GCN is proven to gather contextual information of text. We thus experimented on both of the model in a special setup for both of the datasets.

\section{Selection of Baseline Models} 
\justify
We conducted our experimental performance on two baseline models, GCN and HuBERT. This subchapter describes the reasoning behind choosing these two models. GsCN performs better for gathering textual information due to its benefit for learning knowledge graph, KG \cite{TIAN2022100159} from texts. And thus it can enhance the contextual power of gathering contextual information from speech.
\justify
On the other hand, choosing HuBERT to capture spectral information rather than going for traditional deep learning approaches like CNN and LSTM is that HuBERT is a self-supervised speech model, separately designed for automatic speech recognition downstream tasks not too long ago. And also it is seen that wav2vec models tend to perform better than any other models for the SER tasks. HuBERT is an updated version of the wav2vec2.0 model and thus it can actually overcome the downstream task very accurately. Also as different types of emotion hit for different types of speech information, so capturing both spectral and contextual seemed a suitable task for our research.

\pagebreak
\section{Experiments}
We did two types of experiments in our task, one is traditional deep learning approaches for SER and GCN and HuBERT-based models. The below subsection describes it more.

\subsection{Motivaiton of the experiment}
\justify
The motivation for fusing two different types of architecture, specifically the wav2vec2.0 model and the text2graph model, is to increase the overall accuracy and effectiveness of an emotion recognition system. The fusion of these architectures aims to leverage their respective strengths and enhance the performance of the system.

\justify
The wav2vec2.0 model and the text2graph model are being combined in order to increase the accuracy of the emotion recognition system. The wav2vec2.0 model was created primarily for processing and comprehending voice signals in automatic speech recognition systems. The text2graph model, on the other hand, focuses on transforming textual data into graph representations, making it possible to model links and dependencies between various textual pieces.

\justify
By combining both models, the system may take use of the wav2vec2.0 model's skills in voice signal analysis and feature extraction as well as the text2graph model's aptitude for identifying linkages and dependencies within textual data. A system for emotion recognition that is more complete and reliable can be created by combining these two methods.

\justify
Additionally, the integration of various architectures attempts to improve the system's overall effectiveness in recognizing emotions. The system can have a more comprehensive grasp of emotions communicated in speech and text by combining the advantages of the two models. This can result in more accurate and consistent outcomes by improving the ability to recognize and interpret emotions.

\justify
To improve the precision and efficiency of the emotion identification system, the wav2vec2.0 model and the text2graph model were combined. The combination of these architectures enables a thorough analysis of both speech signals and textual data, utilizing each architecture's advantages to raise the system's total performance.

\justify
Another motivation for our work to have two types of experiment is getting the multimodality of speech data, one is in textual form and another is in spectral form. Multimodal speech emotion recognition are proven to have better performance than the single modality approach.

\section{Experiment Setting and Pre-processing}
\textbf{Experiment I}
\justify
In the first phase of the experiment, we did work with HuBERT pre-trained model for the downstream task of speech emotion recognition, Though it is a pre-trained model to manage with a huge sized dataset we had to take the help of the computational resources. The environment we worked on was having the below specification
\begin{itemize}
    \item CPU: Intel Core i9 12th gen
    \item RAM: 64 GB DDR5
    \item GPU: NVIDIA GeForce RTX 3090
    \item VRAM: 24+24 GB
\end{itemize}

\subsection{Preprocessing I}
Preprocessing of the pipeline contains two part for two datasets.
\begin{itemize}
    \item IEMOCAP dataset contains 5 sessions of acted speech and video. 
    \item We had to set the locations of each session and conduct our experiment for each of the sessions.
    \item As direct audio data are streamed to the pipeline, 5 fold test was performed. 
    \item RAVDESS dataset doesn't contain multiple sessions unlike IEMOCAP. So the full dataset was serially streamed to the HuBERT pipeline. And also RAVDESS is quite smaller than the IEMOCAP, so no sessions were needed.
\end{itemize}

\justify
\clearpage
\textbf{Experiment II}
\justify
Now in the second part of the experiment, we worked with the GCN pipeline to have the stream of text to graph data. 
\justify
And the setting environment is our personal laptop configuration for experimenting with the GCN model. We worked with this experiment earlier HuBERT so the setup was in our personal laptop.

\begin{itemize}
    \item CPU: Intel Core i5 7th gen
    \item RAM: 16 GB DDR4
    \item GPU: Nvidia GeForce GTX 1050
    \item VRAM: 4 GB
\end{itemize}
\subsection{Preprocessing II}
As GCN is there for capturing the contextual information from speech texts, we need not to had much computational power like processing speech data or so. For both of the experiments, we used CUDA to optimize our computational power.

\justify
Now about the preprocessing task, we had to preprocess both of the datasets for GCN and HuBERT pipeline, as GCN is there to work with textual data, we had to prepare transcripts for that. Below is some of the sample of the textual transcripts for the IEMOCAP dataset.

\begin{figure}[H]
    \centering
    \includegraphics[scale=0.355]
    {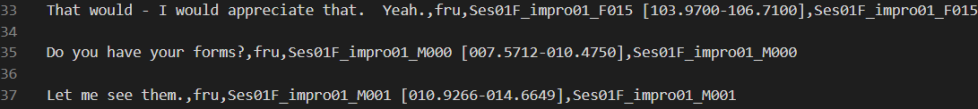}
      \caption {\emph{IEMCOAP text feature generation from speech}}
\end{figure}
And we had to perform the 5 session transcripts for each of the sessions, the transcripts are like the below figure,
\begin{figure}[H]
    \centering
    \includegraphics[scale=0.355]
    {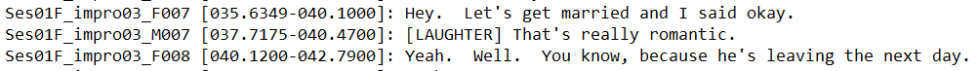}
      \caption {\emph{IEMOCAP transcripts for each sessions}}
\end{figure}

\subsection{Evaluation Metrics and Parameters}
In this subsection, we will discuss about the evaluation metrics we used for our HuBERT pipeline and number of parameters that are generated from our pipeline.

\subsubsection{Dropout}
\justify
Dropout basically refers to a regularization technique in machine learning model that prevents the model to be overfitted and improves the generalization of the model. As the name suggests it basically randomly drops some of the internal nodes in between the network during training iteration. This prevents the model from relying on underlying some nodes and thus it prevents overfitting. 

\justify
Now in the context of speech emotion recognition tasks and specifically for our task where comes the idea of using parallel architecture sequentially, dropout is ineffective in these cases. The goal of the simultaneous training of many neural networks with various designs is to build an ensemble of models. The final forecast is frequently an accumulation or mixture of the distinct predictions made by each model in the ensemble. Dropout, which during training arbitrarily ignores or eliminates layer outputs, may not be required or advantageous in this case because the ensemble already offers a type of regularization through diversity.

\justify
The diversity of the individual models helps the parallel model since each one may identify various elements or patterns in the data. By reducing model bias and raising overall prediction accuracy, this variety helps. And so the parallel approach takes advantage of the inherent diversity among the ensemble members in order to introduce randomness and regularization as opposed to depending on dropout.

\subsubsection{Accuracy matrices}
In the context of our work, some of the accuracy matrices that are frequently used in this domain are:

\begin{itemize}
    \item \textbf{Recognition Accuracy:} This statistic assesses how well the voice emotion identification system distinguishes and categorizes emotions from speech signals. Typically, it is calculated as the proportion of correctly categorized emotion samples among all samples.
    \item \textbf{Confusion Matrix:} The count of predicted emotion labels in comparison to real emotion labels is displayed in the confusion matrix, which is a tabular form. By showing the number of true positives, true negatives, false positives, and false negatives for each emotion class, it gives precise information on the system's performance. Numerous accuracy metrics, including as precision, recall, and F1-score for each emotion category, can be generated from the confusion matrix.
    \item \textbf{Receiver Operating Characteristic (ROC) Curve and Area Under the Curve (AUC):} The trade-off between the true positive rate (sensitivity) and the false positive rate (specificity) for various categorization thresholds is represented graphically by the ROC curve. It is employed to assess a binary classifier's performance. The AUC measures performance across all potential categorization criteria and stands for the area under the ROC curve. Better discriminating between various emotion groups is indicated by a higher AUC.
    \item \textbf{Mean Accuracy:} Measuring the mean of the recognition accuracies achieved for each individual emotion class yields the average accuracy. It gives a general assessment of the system's effectiveness across many emotion categories. 
    \item \textbf{Precision, Recall, and F1-Score:} These metrics are frequently employed in multi-class classification tasks to assess how well each emotion class is performing. Out of all cases projected as positive for a given emotion class, precision is the percentage of correctly predicted positive instances. The percentage of accurately predicted positive instances out of all actual positive instances is measured by the recall, also known as sensitivity. The F1-score, which provides a balanced assessment of the model's performance on a certain emotion class, is the harmonic mean of precision and recall.
    For our work, we used the below F-1 score to measure our accuracy for the proposed model,
    $
    F_{1} = 2\frac{precision.recall}{precision + recall}.
    $
\end{itemize}

\subsubsection{Number of Parameters}
\justify
The number of parameters depends on the size of both of the models and how the input features are augmented with the models. For our GCN architecture, the parameter size was 30K in our experiment. For HuBERT model as this is a pre-trained model the parameter size is unknown to us but we used the base pre-trained model rather than the large model to mitigate the large parameter size and make the model a little bit lightweight.

\subsubsection{Cross Validation and Randomness}
\justify
K-fold cross-validation internally happens in the HuBERT pre-trained architecture so we didn't have to think about that but for the GCN case, we did perform the 5-fold test for the IEMOCAP dataset as that contains 5 sessions of the dataset. Also, the design process for the cross-validation was this way:

\begin{itemize}
    \item \textbf{Splitting the dataset:} Training, validation, and testing sets are commonly used to divide up the speech emotion recognition dataset. To provide a robust evaluation of the model's performance, cross-validation approaches, such as k-fold cross-validation, can be used.
    \item \textbf{Training and validation iterations:} During cross-validation, the model is trained on a subset of the data (training set) and evaluated on another subset (validation set). This process is repeated for different subsets of the data to obtain more reliable performance estimates.
    \item \textbf{Performance metrics:} The measurement of numerous performance metrics, including accuracy, precision, recall, and F1-score, which offer perceptions of the model's performance across various folds, is made possible through cross-validation.
\end{itemize}

\justify
It's important to note that the specific details of cross-validation and randomness techniques can vary depending on the specific approach and research work in this domain.

\chapter{Results and Discussions} 

\label{result} 

\lhead{Chapter \ref{result}. \emph{Experiment Results and Discussions}} 

We conducted our experiment both on GCN and HuBERT architecture and our chosen datasets were RAVDESS and IEMOCAP. For both of the cases, we found comparable performance with the SOTA.

\begin{table}[]
\begin{tabular}{|c|cc|cc|c|}
\hline
\textbf{Model/Dataset}                 & \multicolumn{2}{c|}{\textbf{IEMOCAP}}                 & \multicolumn{2}{c|}{\textbf{RAVDESS}}                 & \textbf{Inference Time}                  \\ \hline
\multirow{2}{*}{\textbf{GCN}}          & \multicolumn{2}{c|}{\multirow{2}{*}{60.47\%}}         & \multicolumn{2}{c|}{\multirow{2}{*}{80.34\%}}         & \multirow{2}{*}{$\sim$35, $\sim$25 mins} \\
                                       & \multicolumn{2}{c|}{}                                 & \multicolumn{2}{c|}{}                                 &                                          \\ \hline
\multirow{2}{*}{\textbf{HuBERT}}       & \multicolumn{2}{c|}{\multirow{2}{*}{62.58\%}}         & \multicolumn{2}{c|}{\multirow{2}{*}{78.01\%}}         & \multirow{2}{*}{$\sim$43, $\sim$23 mins} \\
                                       & \multicolumn{2}{c|}{}                                 & \multicolumn{2}{c|}{}                                 &                                          \\ \hline
\multirow{2}{*}{\textbf{Score (Ours)}} & \multicolumn{1}{c|}{\textbf{FLF}} & \textbf{SLF(max)} & \multicolumn{1}{c|}{\textbf{FLF}} & \textbf{SLF(max)} & \multirow{2}{*}{---}                     \\ \cline{2-5}
                                       & \multicolumn{1}{c|}{-}            & 62.58\%           & \multicolumn{1}{c|}{-}            & 80.34\%           &                                          \\ \hline
\end{tabular}
\caption{Accuracy Table; Feature Level Fusion (FLF) and Score Level Fusion (SLF)}
\end{table}

    \justify
    For both cases, we saw some differences in the conversational dataset and the non-conversational dataset. IEMOCAP being the conversational dataset performs better on the wave-to-vector model, that is HuBERT, and RAVDESS not being the conversational dataset performs better on the Graph representational model that is GCN.

    \justify
    We compared our result with the SOTA performance. The formulated table is given below for the IEMOCAP dataset. Our model outperforms 3 of the available benchmarks:

    \begin{table}[ht]
    \centering
\begin{tabular}{|c|c|}
\hline
\textbf{SER Models}                                       & \textbf{Accuracy (ACC\%)} \\ \hline
Attn-BLSTM 2016\cite{huang2016attention} & 59.33                     \\ \hline
RNN 2017\cite{mirsamadi2017automatic}    & 63.50                     \\ \hline
CRNN 2018\cite{luo2018investigation}     & 63.98                     \\ \hline
LSTM 2019\cite{latif2019direct}          & 58.72                     \\ \hline
CNN-LSTM 2019\cite{latif2019direct}      & 59.23                     \\ \hline
\textbf{Ours}                                             & \textbf{62.58}            \\ \hline
\end{tabular}
\caption{SER results and comparison on the IEMOCAP dataset}
\end{table}

     For the RAVDESS dataset:

    \begin{table}[ht]
    \centering
\begin{tabular}{|c|c|}
\hline
\textbf{SER Models}                                     & \textbf{Accuracy (ACC\%)} \\ \hline
SVM\cite{bhavan2019bagged}             & 79.10                     \\ \hline
AlexNet 2021\cite{luna2021multimodal}  & 61.67                     \\ \hline
CNN-14 2021\cite{luna2021multimodal}   & 76.58                     \\ \hline
TIM-Net 2022\cite{ye2023temporal}      & 92.08                     \\ \hline
Wav2Vec2.0 2021\cite{luna2021proposal} & 81.82                     \\ \hline
\textbf{Ours}                                           & \textbf{80.34}            \\ \hline
\end{tabular}
\caption{SER results and comparison on the RAVDESS dataset}
\end{table}

    \justify
    Weighted accuracy (WA) is an evaluation metric used in classification tasks that considers the varying importance or distribution of different classes or data points in a dataset. It takes into account the fact that some classes or data points may have a higher significance or imbalance in representation. By assigning predetermined weights to each class or data point, weighted accuracy provides a more accurate measure of performance that reflects the relative importance of different elements in the dataset. This metric is particularly useful when dealing with imbalanced datasets or when certain classes or data points carry more weight in the evaluation. Overall, weighted accuracy offers a more nuanced assessment of classification models by incorporating varying degrees of importance within the dataset.

    \justify
    Surprisingly in the existing literatures, Wav2Vec2.0 model showed better performance among all the models, for our case HuBERT also showed comparative performance for RAVDESS.

    \justify
    For the IEMOCAP dataset in the existing literature CRNN showed the best performance, for our case HuBERT performed better than GCN. 

    \justify
    Nonetheless, increasing accuracy was not the main concern of our work rather capturing both long-term contextual and spectral information at the same time to make our pipeline more robust in the SER field was the actual contribution.

    \clearpage
    \justify
    \section{HuBERT-RAVDESS experiment}
    \justify
    Iteration vs. Training Loss
    \begin{figure}[!ht]
    \centering
    \includegraphics[scale=0.9]
    {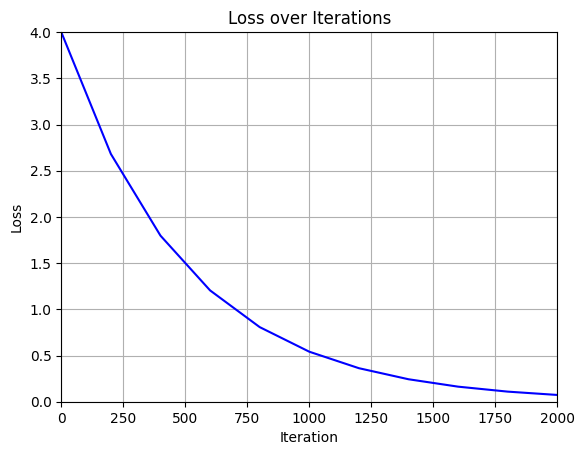}
      \caption {\emph{RADVDESS HuBERT iteration for 4 specific emotion classes}}
    \end{figure}
    \justify
    In this figure denoted graph is actually the loss and iteration graph for ravdess-hubert. Initially, when the iteration was between 0 to 200 we can see the loss was greater and then eventually the loss became less reaching the iteration of 2000. It is visible that the loss is tends to 0 for more number of iterations.

    \clearpage
    \begin{figure}[!ht]
    \centering
    \includegraphics[scale=0.9]
    {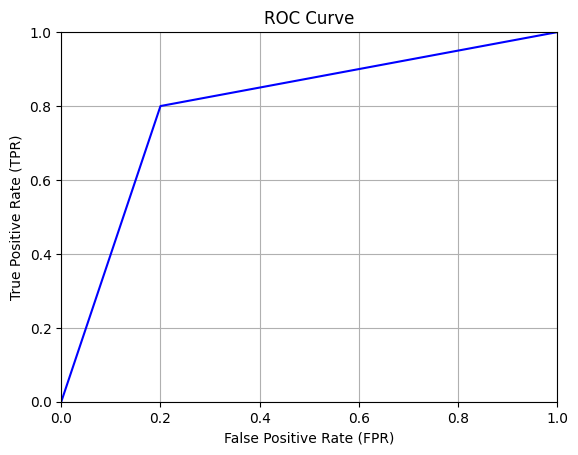}
      \caption {\emph{RADVDESS HuBERT ROC curve}}
    \end{figure}
    \justify
    This figure shows the True Positive Rate (TPR) vs False Positive Rate (FPR) graph, which is also known as the ROC curve, denoting the performance of our classifier. We can see that initially, the TPR is high, indicating the total of positive instances that are actually positive.  

    \clearpage
    \justify
    Epoch vs. Training Accuracy
    \begin{figure}[!ht]
    \centering
    \includegraphics[scale=0.9]
    {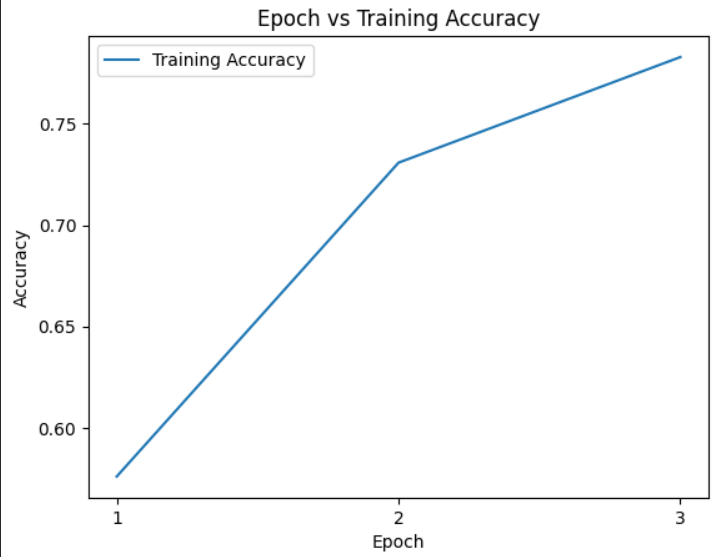}
      \caption {\emph{RADVDESS HuBERT Training testing for 4 specific emotion classes}}
    \end{figure}
    \justify
    In this figure, we can see that the epoch level reaching to 2, then the accuracy took a turn and went for less increasing way than before, so we had to early stop to get a comparable accuracy. That's why 3 epochs seemed right to us.
    \clearpage
    
    \justify
    Confusion Matrix for HuBERT-RAVDESS experiment
    \begin{figure}[!ht]
    \centering
    \includegraphics[scale=0.9]
    {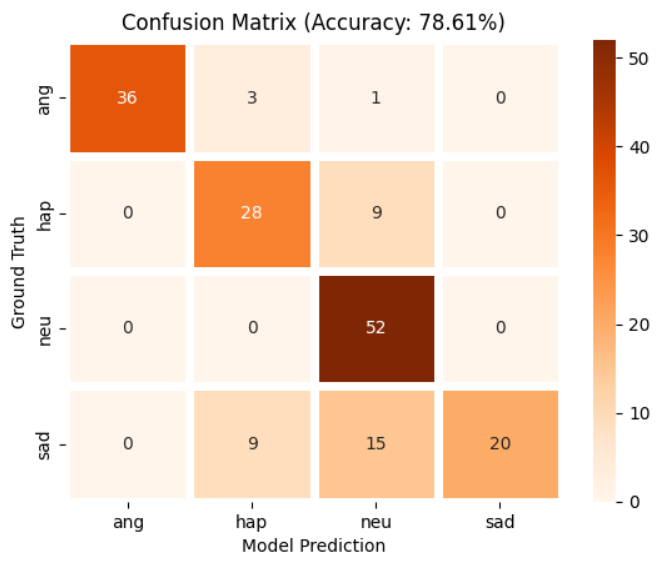}
      \caption {\emph{RADVDESS HuBERT Confusion Matrix for 4 specific emotion classes}}
    \end{figure}

    \justify
    \textbf{Descriptions:} 
    \begin{itemize}
        \item 4 class emotions; Angry, Happy, Neutral and Sad.
        \item True Positive (TP): The model predicts positive, and the actual label is also positive. False Positive (FP): The model predicts positive, but the actual label is negative. False Negative (FN): The model predicts a negative, but the actual label is positive. True Negative (TN): The model predicts a negative, and the actual label is also negative.
        \item For angry TP is 36 which means 36\% of data are True Positive and accurately classified. The rest of the 4\% are False Negative. There are no False Positive for this emotion class.
        \item For happy class, 26 are TP, 12 are FP, and 9 are False Negative.
        \item  For the neutral case, the highest TP is recorded, which means the model performs very well for the neutral emotion class.
        \item However, overall accuracy  of the model is 78.61\%.
    \end{itemize}



    \section{GCN-IEMOCAP experiment}
    \justify
    Our conducted experiment on IEMOCAP dataset for GCN gave us the following results having compared with the existing work.
    \begin{itemize}
        \item epoch: 45
        \item epoch time: 3.480687 sec
        \item loss training: 0.775471
        \item Early Stopping Counter: 10 out of 10
        \item Accuracy train: 0.673471
        \item Accuracy test: 0.627498
        \item Average train acc: 0.682884
        \item Average test acc: 0.624793
    \end{itemize}

    \justify
    Reasonings behind the experimented parameters:
    \begin{itemize}
        \item We experimented and chose to stay on 45 epochs as given more epochs our model was ovcerfitting.
        \item As the IEMOCAP dataset was 5 fold splitting considering the 5 sessions, we had to also split and use the first 4 sessions as our training sessions and the last session as our testing session.  
        \item We had to use an early stopping mechanism having a counter of 10, which means that our model was not improving past 10.
        \item The average training testing accuracy of the model gives an idea of how our model performs in general.
    \end{itemize}

\justify

\chapter{Conclusion and Future Work} 

\label{Conclusion} 

\lhead{Chapter \ref{Conclusion}. \emph{Conclusion and Future Work}}

\section{Conclusion}
    \justify
    Techniques for multimodal speech emotion detection that accept input from both speech and text usually outperform those that employ only one modality because they can make use of the complimentary information each modality offers. Both speech and writing may offer a variety of clues about the speaker's emotional state, making them both valuable sources of information about a speaker's state of mind. For instance, the text has semantic and grammatical indicators that can also give hints about the speaker's emotional state, whereas speech signals contain prosodic cues like pitch, intensity, and length that might transmit emotional information. Multimodal techniques frequently perform better than single-modality approaches because they may leverage the advantages of both modalities' information by integrating them.

    \justify
    Multimodal speech emotion recognition methodologies may be applied in a variety of ways. One approach is to employ a single model that learns to extract characteristics from both voice and text through the utilization of input from both modalities. Using different models for voice and text and fusing their output is an alternative strategy.

    \justify
    Overall, compared to single-modality approaches, multimodal speech emotion identification algorithms have the potential to produce more reliable and accurate findings, especially when one modality is noisy or absent. It's crucial to keep in mind, too, that multimodal approaches can also increase complexity and call for more data and processing power to develop and implement.

    \justify
    To imply our work, it can be used for our case combining the output of speech and textual data to increase the integrity of classification but that might be a complex process as have to process both speech and textual data and build different architectures to support those.

    \justify
    To conclude the idea of using a parallel GCN and HuBERT architecture is that first of all, we will be able to represent contextual features conveniently and robustly while processing with GCN, and at the same time, we will be using the fundamental theory of spectral Graph Convolutional Network, working with the Fourier domain to extract important emotional features. And the second idea of using HuBERT is its ability to capture long-range spectral dependencies in the speech signal by passing MFCC spectrograms to the network. One key idea to using parallelism of both of the networks is influenced by GoogleNet due to the "inception modules," which are designed to capture both local and global features from the input image. The inception modules consist of a series of convolutional and pooling layers that are applied in parallel, with the output of each layer being concatenated and fed into the next layer. This allows the network to learn from different types of features simultaneously and to capture more fine-grained information from the input image like MFCC spectrograms. Though there hasn't been much work on graph architecture for SER, we are hoping to implement some experiments toward our understanding and make some results.

\section{Future Work}
In future research, there is an opportunity to further explore the potential of combining speech and textual data to enhance the integrity of speech emotion classification. However, this endeavor entails addressing the complexities involved in effectively processing and integrating both modalities. Developing different architectures to accommodate and extract meaningful features from speech and text inputs will be essential. Additionally, investigating parallel models, such as the parallel GCN and HuBERT architecture, holds promise. This approach allows for the simultaneous representation of contextual features and the extraction of long-range spectral dependencies, leading to a more comprehensive analysis. Despite the limited research on graph architectures for speech emotion recognition, conducting experiments and analyzing the results can contribute to the advancement and understanding of this field. It is through such explorations that the field of multimodal speech emotion recognition can continue to progress, enabling more accurate and nuanced assessments of human emotions.




\addtocontents{toc}{\vspace{2em}} 

\appendix 


\chapter{Appendix} 

\label{AppendixA} 

\lhead{Appendix \emph{A}} 


\section{Machine Learning Models}
    Speech Emotion Recognition (SER) using traditional machine learning approaches involves the application of techniques such as Support Vector Machines (SVM), Hidden Markov Models (HMM), and Gaussian Mixture Models (GMM) for emotion classification. These models utilize handcrafted features extracted from speech signals to train classifiers. SVM separates different emotion classes using a hyperplane, while HMM models capture temporal dependencies and transitions between emotional states. GMM represents speech data as a mixture of Gaussian distributions. Although these traditional machine learning approaches have been used successfully in SER, they rely heavily on manually engineered features and may struggle with capturing complex emotional expressions and long-term dependencies in speech.
    \subsection{SER using Support Vector Machine}
    \justify
   A Support Vector Machine (SVM) is a machine learning algorithm used in speech emotion recognition tasks to classify the emotional state of individuals based on their speech signals. It involves extracting relevant features from the speech data, normalizing them, and training the SVM model using labeled examples. The SVM learns to find an optimal hyperplane in a high-dimensional feature space that maximally separates different emotional classes, enabling accurate classification of unseen speech samples. By leveraging kernel functions and fine-tuning parameters, SVMs offer a robust and efficient approach for identifying emotions from speech signals.

\subsection{SER using Hidden Markov Model}
    \justify
     
    A Hidden Markov Model (HMM) is a statistical model commonly used for speech emotion recognition tasks. In this context, an HMM represents the underlying emotional states as hidden states and the observed speech features as emitted symbols. The HMM consists of transition probabilities between the hidden states, emission probabilities that link the hidden states to the observed speech features, and initial state probabilities. The model is trained using labeled speech data, where the emotional state sequences are known. During training, the HMM estimates the parameters that maximize the likelihood of the observed speech data given the emotional state sequences. Once trained, the HMM can be used to predict the emotional state of unseen speech samples by finding the most probable sequence of hidden states given the observed speech features, using algorithms like the Viterbi algorithm. HMMs provide a flexible and probabilistic framework for modeling temporal dependencies in speech and are widely used in speech emotion recognition tasks.

\subsection{SER using Gaussian Mixture Model}
    \justify
    
    A Gaussian Mixture Model (GMM) is a statistical model frequently utilized in speech emotion recognition tasks. In this context, a GMM represents the probability distribution of speech features within each emotional state using a combination of Gaussian distributions. Each emotional state is associated with a mixture component, and the model parameters consist of the mean, covariance, and weight of each Gaussian distribution. During training, the GMM is fitted to the labeled speech data by estimating the model parameters that maximize the likelihood of the observed features. To predict the emotional state of unseen speech samples, the GMM computes the likelihood of the observed features under each mixture component and assigns the sample to the emotional state associated with the component with the highest likelihood. GMMs provide a flexible and powerful framework for modeling the complex distribution of speech features and are commonly employed in speech emotion recognition to capture the inherent variability within different emotional states.

\section{Deep Learning models}
    \justify
    Generic Deep learning approaches basically consist of a bunch of hidden layers, including input layer and output layers to correctly identify weights and features from the network and successfully classify them. Some old traditional DL approaches were using restricted Boltzmann machine architecture (RBM), and Deep Belief Network Architecture (DBN) consists of a bunch of RBMs, The more improved version of these two later came named Deep Boltzmann Machine Architecture. The benefit of DBM was to learning ability was fast and representation was effective due to layer-wise pre-training.

\subsection{SER using Deep autoencoder}
    The autoencoder technique contributes to having an encoder and decoder in the network, where the encoder is used to capture hidden feature information of input speech data, and the decoder is used to reduce the error percentage of the network.
    \begin{figure}[!hbt]
    \centering
    \includegraphics[scale=0.7]
    {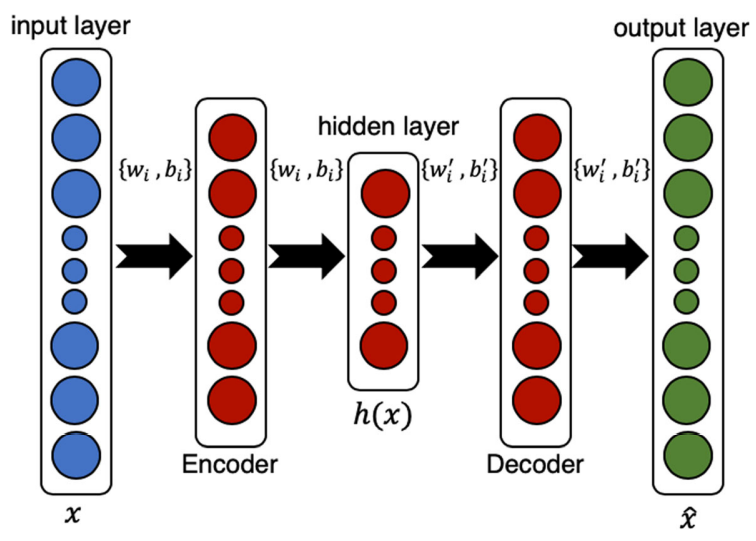}
      \caption {\emph{Deep Autoencoder Architecture\cite{jahangir2021deep}}}
    \end{figure}

\subsection{SER using Convolutional Neural Network}
    CNN model general case having three different layers (convolutional, pooling and fully connected layer) successfully capture the class difference of classification problem, where the main layer convolutional comprises the neurons that capture the image spectrograms of speech with aid of a linear filter. 
    \begin{figure}[!hbt]
    \centering
    \includegraphics[scale=0.7]
    {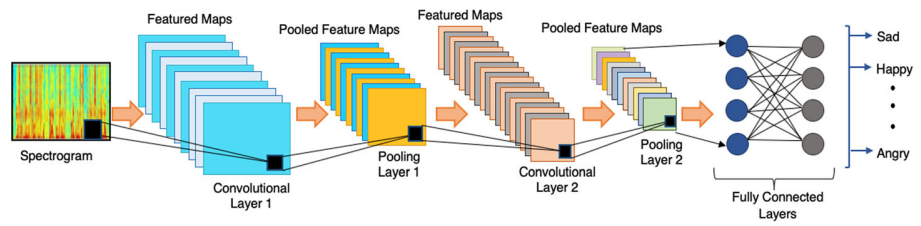}
      \caption {\emph{Convolutional Neural Network Architecture\cite{chauhan2021speech}}}
    \end{figure}

    Then pooling layer does different types of pooling mechanisms including average pooling, stochastic pooling, average pooling, etc. Recently, max pooling showed better performance, hence max pooling technique is now broadly used for the pooling layer due to its structural rigidity of slight variations in speech utterances.

    Then comes the fully connected layer after many convolutional and pooling layers. Fully connected layers connected with classifiers take the master feature vector for classification then successfully classify emotional expressions.

\subsection{SER using Recurrent Neural Network}
    Speech can be represented as a 2-D graph of frequency and time. That's why speech databases are also known as time series data. Sequential information is important hence. Temporal features thus can be captured by sequential representation understanding models like RNN. Different variations of RNN are now used frequently like Long Short Term Memory (LSTM) for capturing long-term temporal dependencies of speech. Though LSTM proves to provide better performance for speech, it eventually contributes to the complexity and exploding gradient problems. 
    \begin{figure}[!hbt]
    \centering
    \includegraphics[scale=0.5]
    {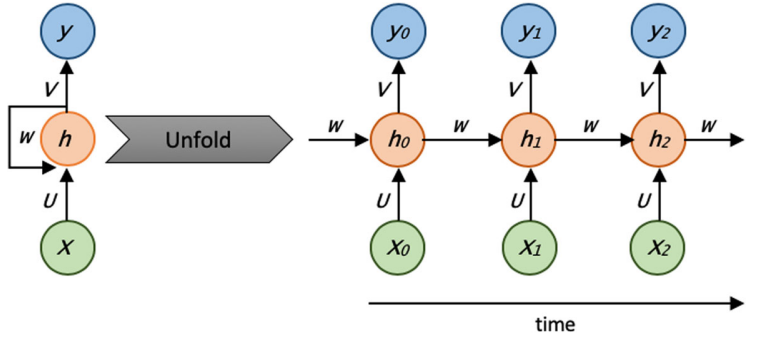}
      \caption {\emph{Recurrent Neural Network Architecture\cite{han2021speech}}}
    \end{figure}

\section{SER using Evolution of Deep Learning Techniques}
    Over the years, DL techniques for SER improved hugely and it can be broadly categorized into three, Generative, Discriminative, and Hybrid. In recent times due to the complexity of Discriminative models, Hybrid models thus perform more effectively than others. Taxonomy is given below for the evolution of models of SER.
    
\subsection{SER using Graph Convolution Network(GCN)}
    \justify
    
    A Graph Convolution Network (GCN) model for speech emotion recognition is a deep learning architecture designed to capture the relationships between speech features as a graph structure. In this context, the nodes of the graph represent speech features or segments, while the edges denote the connections or relationships between them. The GCN model incorporates graph convolutional layers, which enable information propagation across the graph by aggregating neighboring nodes' information. By leveraging these graph convolutions, the model can learn and extract high-level representations that capture both local and global dependencies in the speech data. The GCN model is trained using labeled speech samples, where the emotional states are known. During training, the model learns to map the graph-based representations of speech features to the corresponding emotional states. This approach allows the GCN model to effectively capture and utilize the inherent structural relationships in the speech data, making it suitable for speech emotion recognition tasks.

\subsection{SER using Hidden Unit BERT (HuBERT) model}

    \justify

   HuBERT (Hidden Unit Bidirectional Encoder Representations from Transformers) is a self-supervised learning method for speech representation that focuses on masked prediction of hidden units. It utilizes a deep neural network architecture to learn meaningful representations from unlabeled speech data. In the training process, a portion of the hidden units is randomly masked, and the model is tasked with predicting the masked units based on the remaining information. By learning to fill in the missing information, HuBERT captures intricate patterns and structures in the speech data. This self-supervised approach enables the model to acquire high-level representations without the need for manual annotations. HuBERT has shown promising results in various speech-related tasks, including speech emotion recognition, by producing discriminative and informative representations that enhance the performance of downstream tasks.

\subsection{Multimodal Speech Emotion Recognition and Classification Using Convolutional Neural Network Techniques}
    \justify
    Multimodal Speech Emotion Recognition and Classification using Convolutional Neural Network (CNN) techniques refers to an approach that leverages multiple modalities, such as speech and visual cues, for accurately identifying and categorizing emotions in speech. The system combines CNN architectures with multimodal data, allowing the model to learn complex patterns and representations from both the audio and visual features. The speech data is typically transformed into spectrograms or other suitable representations, while visual information may include facial expressions or gestures. The CNN model processes the multimodal inputs through convolutional layers to extract relevant features and capture the spatial and temporal relationships between them. By jointly analyzing audio and visual cues, this approach enhances the recognition and classification accuracy of emotions in speech, leading to more robust and comprehensive models for multimodal speech emotion analysis.

\addtocontents{toc}{\vspace{2em}} 

\backmatter


\label{Bibliography}

\lhead{\emph{Bibliography}} 

\bibliographystyle{IEEEtran} 

\bibliography{Bibliography} 

\begin{thebibliography}{10}
\providecommand{\url}[1]{#1}
\csname url@samestyle\endcsname
\providecommand{\newblock}{\relax}
\providecommand{\bibinfo}[2]{#2}
\providecommand{\BIBentrySTDinterwordspacing}{\spaceskip=0pt\relax}
\providecommand{\BIBentryALTinterwordstretchfactor}{4}
\providecommand{\BIBentryALTinterwordspacing}{\spaceskip=\fontdimen2\font plus
\BIBentryALTinterwordstretchfactor\fontdimen3\font minus
  \fontdimen4\font\relax}
\providecommand{\BIBforeignlanguage}[2]{{%
\expandafter\ifx\csname l@#1\endcsname\relax
\typeout{** WARNING: IEEEtran.bst: No hyphenation pattern has been}%
\typeout{** loaded for the language `#1'. Using the pattern for}%
\typeout{** the default language instead.}%
\else
\language=\csname l@#1\endcsname
\fi
#2}}
\providecommand{\BIBdecl}{\relax}
\BIBdecl

\bibitem{jahangir2021deep}
R.~Jahangir, Y.~W. Teh, F.~Hanif, and G.~Mujtaba, ``Deep learning approaches
  for speech emotion recognition: State of the art and research challenges,''
  \emph{Multimedia Tools and Applications}, vol.~80, no.~16, pp.
  23\,745--23\,812, 2021.

\bibitem{jain2020speech}
M.~Jain, S.~Narayan, P.~Balaji, A.~Bhowmick, R.~K. Muthu \emph{et~al.},
  ``Speech emotion recognition using support vector machine,'' \emph{arXiv
  preprint arXiv:2002.07590}, 2020.

\bibitem{Cheng2012/08}
\BIBentryALTinterwordspacing
X.~Cheng and Q.~Duan, ``Speech emotion recognition using gaussian mixture
  model,'' in \emph{Proceedings of the 2012 International Conference on
  Computer Application and System Modeling (ICCASM 2012)}.\hskip 1em plus 0.5em
  minus 0.4em\relax Atlantis Press, 2012/08, pp. 1222--1225. [Online].
  Available: \url{https://doi.org/10.2991/iccasm.2012.311}
\BIBentrySTDinterwordspacing

\bibitem{chauhan2021speech}
K.~Chauhan, K.~K. Sharma, and T.~Varma, ``Speech emotion recognition using
  convolution neural networks,'' in \emph{2021 international conference on
  artificial intelligence and smart systems (ICAIS)}.\hskip 1em plus 0.5em
  minus 0.4em\relax IEEE, 2021, pp. 1176--1181.

\bibitem{han2021speech}
S.~Han, F.~Leng, and Z.~Jin, ``Speech emotion recognition with a
  resnet-cnn-transformer parallel neural network,'' in \emph{2021 International
  Conference on Communications, Information System and Computer Engineering
  (CISCE)}.\hskip 1em plus 0.5em minus 0.4em\relax IEEE, 2021, pp. 803--807.

\bibitem{issa2020speech}
D.~Issa, M.~F. Demirci, and A.~Yazici, ``Speech emotion recognition with deep
  convolutional neural networks,'' \emph{Biomedical Signal Processing and
  Control}, vol.~59, p. 101894, 2020.

\bibitem{amiriparian2021impact}
S.~Amiriparian, A.~Sokolov, I.~Aslan, L.~Christ, M.~Gerczuk, T.~H{\"u}bner,
  D.~Lamanov, M.~Milling, S.~Ottl, I.~Poduremennykh \emph{et~al.}, ``On the
  impact of word error rate on acoustic-linguistic speech emotion recognition:
  an update for the deep learning era,'' \emph{arXiv preprint
  arXiv:2104.10121}, 2021.

\bibitem{scheidwasser2022serab}
N.~Scheidwasser-Clow, M.~Kegler, P.~Beckmann, and M.~Cernak, ``Serab: A
  multi-lingual benchmark for speech emotion recognition,'' in \emph{ICASSP
  2022-2022 IEEE International Conference on Acoustics, Speech and Signal
  Processing (ICASSP)}.\hskip 1em plus 0.5em minus 0.4em\relax IEEE, 2022, pp.
  7697--7701.

\bibitem{wang2021learning}
Y.~Wang, G.~Shen, Y.~Xu, J.~Li, and Z.~Zhao, ``Learning mutual correlation in
  multimodal transformer for speech emotion recognition.'' in
  \emph{Interspeech}, 2021, pp. 4518--4522.

\bibitem{ho2020multimodal}
N.-H. Ho, H.-J. Yang, S.-H. Kim, and G.~Lee, ``Multimodal approach of speech
  emotion recognition using multi-level multi-head fusion attention-based
  recurrent neural network,'' \emph{IEEE Access}, vol.~8, pp. 61\,672--61\,686,
  2020.

\bibitem{wang2021novel}
X.~Wang, M.~Wang, W.~Qi, W.~Su, X.~Wang, and H.~Zhou, ``A novel end-to-end
  speech emotion recognition network with stacked transformer layers,'' in
  \emph{ICASSP 2021-2021 IEEE International Conference on Acoustics, Speech and
  Signal Processing (ICASSP)}.\hskip 1em plus 0.5em minus 0.4em\relax IEEE,
  2021, pp. 6289--6293.

\bibitem{shirian2021compact}
A.~Shirian and T.~Guha, ``Compact graph architecture for speech emotion
  recognition,'' in \emph{ICASSP 2021-2021 IEEE International Conference on
  Acoustics, Speech and Signal Processing (ICASSP)}.\hskip 1em plus 0.5em minus
  0.4em\relax IEEE, 2021, pp. 6284--6288.

\bibitem{nwe2003speech}
T.~L. Nwe, S.~W. Foo, and L.~C. De~Silva, ``Speech emotion recognition using
  hidden markov models,'' \emph{Speech communication}, vol.~41, no.~4, pp.
  603--623, 2003.

\bibitem{bhavan2020deep}
A.~Bhavan, M.~Sharma, M.~Piplani, P.~Chauhan, Hitkul, and R.~R. Shah, ``Deep
  learning approaches for speech emotion recognition,'' \emph{Deep
  learning-based approaches for sentiment analysis}, pp. 259--289, 2020.

\bibitem{li2021speech}
D.~Li, J.~Liu, Z.~Yang, L.~Sun, and Z.~Wang, ``Speech emotion recognition using
  recurrent neural networks with directional self-attention,'' \emph{Expert
  Systems with Applications}, vol. 173, p. 114683, 2021.

\bibitem{heusser2019bimodal}
V.~Heusser, N.~Freymuth, S.~Constantin, and A.~Waibel, ``Bimodal speech emotion
  recognition using pre-trained language models,'' \emph{arXiv preprint
  arXiv:1912.02610}, 2019.

\bibitem{zhao2019speech}
J.~Zhao, X.~Mao, and L.~Chen, ``Speech emotion recognition using deep 1d \& 2d
  cnn lstm networks,'' \emph{Biomedical signal processing and control},
  vol.~47, pp. 312--323, 2019.

\bibitem{huang2014speech}
Z.~Huang, M.~Dong, Q.~Mao, and Y.~Zhan, ``Speech emotion recognition using
  cnn,'' in \emph{Proceedings of the 22nd ACM international conference on
  Multimedia}, 2014, pp. 801--804.

\bibitem{christy2020multimodal}
A.~Christy, S.~Vaithyasubramanian, A.~Jesudoss, and M.~A. Praveena,
  ``Multimodal speech emotion recognition and classification using
  convolutional neural network techniques,'' \emph{International Journal of
  Speech Technology}, vol.~23, pp. 381--388, 2020.

\bibitem{busso2008iemocap}
C.~Busso, M.~Bulut, C.-C. Lee, A.~Kazemzadeh, E.~Mower, S.~Kim, J.~N. Chang,
  S.~Lee, and S.~S. Narayanan, ``Iemocap: Interactive emotional dyadic motion
  capture database,'' \emph{Language resources and evaluation}, vol.~42, pp.
  335--359, 2008.

\bibitem{livingstone2018ryerson}
S.~R. Livingstone and F.~A. Russo, ``The ryerson audio-visual database of
  emotional speech and song (ravdess): A dynamic, multimodal set of facial and
  vocal expressions in north american english,'' \emph{PloS one}, vol.~13,
  no.~5, p. e0196391, 2018.

\bibitem{TIAN2022100159}
\BIBentryALTinterwordspacing
L.~Tian, X.~Zhou, Y.-P. Wu, W.-T. Zhou, J.-H. Zhang, and T.-S. Zhang,
  ``Knowledge graph and knowledge reasoning: A systematic review,''
  \emph{Journal of Electronic Science and Technology}, vol.~20, no.~2, p.
  100159, 2022. [Online]. Available:
  \url{https://www.sciencedirect.com/science/article/pii/S1674862X2200012X}
\BIBentrySTDinterwordspacing

\bibitem{huang2016attention}
C.-W. Huang and S.~S. Narayanan, ``Attention assisted discovery of
  sub-utterance structure in speech emotion recognition.'' in
  \emph{Interspeech}, 2016, pp. 1387--1391.

\bibitem{mirsamadi2017automatic}
S.~Mirsamadi, E.~Barsoum, and C.~Zhang, ``Automatic speech emotion recognition
  using recurrent neural networks with local attention,'' in \emph{2017 IEEE
  International conference on acoustics, speech and signal processing
  (ICASSP)}.\hskip 1em plus 0.5em minus 0.4em\relax IEEE, 2017, pp. 2227--2231.

\bibitem{luo2018investigation}
D.~Luo, Y.~Zou, and D.~Huang, ``Investigation on joint representation learning
  for robust feature extraction in speech emotion recognition.'' in
  \emph{Interspeech}, 2018, pp. 152--156.

\bibitem{latif2019direct}
S.~Latif, R.~Rana, S.~Khalifa, R.~Jurdak, and J.~Epps, ``Direct modelling of
  speech emotion from raw speech,'' \emph{arXiv preprint arXiv:1904.03833},
  2019.

\bibitem{bhavan2019bagged}
A.~Bhavan, P.~Chauhan, R.~R. Shah \emph{et~al.}, ``Bagged support vector
  machines for emotion recognition from speech,'' \emph{Knowledge-Based
  Systems}, vol. 184, p. 104886, 2019.

\bibitem{luna2021multimodal}
C.~Luna-Jim{\'e}nez, D.~Griol, Z.~Callejas, R.~Kleinlein, J.~M. Montero, and
  F.~Fern{\'a}ndez-Mart{\'\i}nez, ``Multimodal emotion recognition on ravdess
  dataset using transfer learning,'' \emph{Sensors}, vol.~21, no.~22, p. 7665,
  2021.

\bibitem{ye2023temporal}
J.~Ye, X.-C. Wen, Y.~Wei, Y.~Xu, K.~Liu, and H.~Shan, ``Temporal modeling
  matters: A novel temporal emotional modeling approach for speech emotion
  recognition,'' in \emph{ICASSP 2023-2023 IEEE International Conference on
  Acoustics, Speech and Signal Processing (ICASSP)}.\hskip 1em plus 0.5em minus
  0.4em\relax IEEE, 2023, pp. 1--5.

\bibitem{luna2021proposal}
C.~Luna-Jim{\'e}nez, R.~Kleinlein, D.~Griol, Z.~Callejas, J.~M. Montero, and
  F.~Fern{\'a}ndez-Mart{\'\i}nez, ``A proposal for multimodal emotion
  recognition using aural transformers and action units on ravdess dataset,''
  \emph{Applied Sciences}, vol.~12, no.~1, p. 327, 2021.

\end{thebibliography}

\end{document}